\documentclass[oldversion]{aa}          % Versione senza abstract strutturato
\usepackage{graphicx}%\usepackage{rotating}
\input epsf.sty

\def \grss {GRS 1915+105~}
\def \grs {GRS 1915+105}
\def \trec{$T_{rec}$}
\def \sax {{\it Beppo}SAX}

\begin{document}
%\thesaurus{08.14.1; 08.16.7 PSR~B1509$-$58; 13.25.5)}
\title{The complex behaviour of the microquasar GRS~1915+105 in the $\rho$
class observed with \sax. I: Timing analysis
\thanks{Table A.2 and Table A.3 are available in electronic form at
CDS via anonymous ftp to cdsarc.ustrasbg.fr (130.79.128)
}
}
\author{%ORDINE PRELIMINARE~~
E.~Massaro\inst{1}
\and G.~Ventura\inst{2}
\and F.~Massa\inst{2,3}
\and M.~Feroci\inst{4}
\and T.~Mineo\inst{5}
\and G.~Cusumano\inst{5}
\and P.~Casella\inst{6}
\and T.~Belloni\inst{7}
%\and - - -
\institute{Dipartimento di Fisica, Universit\'a La Sapienza, Piazzale A. Moro 2,
I-00185 Roma, Italy
\and Stazione Astronomica di Vallinfreda, via del Tramonto, Vallinfreda (RM), Italy
\and INFN-Sezione di Roma1 (retired), Roma, Italy
\and INAF, Istituto di Astrofisica Spaziale e Fisica
cosmica di Roma, via del Fosso del Cavaliere 100, I-00113 Roma, Italy
\and INAF, Istituto di Astrofisica Spaziale e Fisica
cosmica di Palermo, via U. La Malfa 153, I-90146 Palermo, Italy
\and Astronomical Institute, University of Amsterdam, The Netherlands
\and INAF, Osservatorio Astronomico di Brera, via E. Bianchi 46, I-23807 Merate, Italy
}}
\offprints{enrico.massaro@uniroma1.it}
\date{Received ....; accepted ....}

\markboth{E. Massaro et al.: The complex behaviour of GRS~1915+105 in the $\rho$ class.
I: Timing analysis}
{E. Massaro et al.: The complex behaviour of GRS~1915+105 in the $\rho$ class.
I: Timing analysis}

\abstract{ \grss was observed by \sax~ for about 10 days in
October 2000. For about 80\% of the time, the source was in the
variability class $\rho$, characterised by a series of recurrent
bursts. We describe the results of the timing
analysis performed on the MECS (1.6--10 keV) and PDS (15--100 keV)
data.
The X-ray count rate from
\grss showed an increasing trend with different characteristics in
the various energy bands: in the bands (1.6--3 keV) and (15--100
keV), it was nearly stable in the first part of the pointing and
increased in a rather short time by about 20\%, while in the
energy range (3--10 keV) the increase
had a smoother trend.\\
Fourier and wavelet analyses detect a variation in the recurrence time
of the bursts, from 45--50 s to about 75 s, which appear well correlated
with the count rate.
From the power distribution of peaks in Fourier periodograms and
wavelet spectra, we distinguished between the {\it regular} and {\it irregular}
variability modes of the $\rho$ class, which are related to variations in 
the count rate in the 3--10 keV range.\\
We identified two components in the burst structure: the slow leading
trail, and the pulse, superimposed on a rather stable level.
Pulses are generally structured in a series of peaks and their
number is related to the regularity modes: the mean number of peaks 
is lower than 2 in the regular mode and increases up to values higher
than 3 in the irregular mode.
We found that the change in the recurrence time of the regular mode
is caused by the slow leading trails, while the duration of
the pulse phase remains far more stable.
The evolution in the mean count rates shows that the time behaviour
of both the leading trail and the baseline level are very similar to those
observed in the 1.6--3 and 15--100 keV ranges, while that of the pulse
follows the peak number.\\
These differences in the time behaviour and count rates at different
energies indicate that the process responsible for the pulses must produce
the strongest emission between 3 and 10 keV, while that associated with
both the leading trail and the baseline dominates at lower and higher energies.
}
\authorrunning{E. Massaro et al.}
\titlerunning{The complex behaviour of GRS~1915+105 in the $\rho$ class}

\maketitle

\section{Introduction}
The galactic microquasar \grss was discovered in 1992 in the hard
X-ray band (Castro-Tirado, Brandt \& Lund 1992).
VLA images of the radio counterpart showed two opposite radio jets moving
at an apparent superluminal velocity, and from their proper motions both a
distance of about 12 kpc and an inclination angle to the line of sight
of about 70$^{\circ}$ were inferred (Mirabel \& Rodriguez 1994).

The optical counterpart was identified by Castro-Tirado et al.
(2001) with a binary system of orbital period about 33
days (Greiner, Cuby \& McCaughrean 2001) containing a black hole
whose estimated mass is  14.0$\pm$4.4 $M_{\odot}$ (Harlaftis \&
Greiner 2004). The X-ray behaviour of \grss is characterised by 
strong variability with very bright and quiet phases. The spectrum
is generally fitted by at least two components: a multi-temperature
disk black body and a power law (possibly with a variable
exponential cut-off) extending up to several hundreds keV. During
the bursts, the main spectral parameters of the thermal component
exhibit significant variations, which were interpreted as representing
the emptying and  refilling of the inner portion of the accretion disk
(Belloni et al. 1997a). Quasi-periodic oscillations were
observed with the PCA experiment onboard RXTE at frequencies up to 
above 100 Hz (Morgan, Remillard \& Greiner 1997;
McClintock \& Remillard  2006) and a strong signal around 67 Hz, 
also studied by Belloni, M\'ndez \& S\'anchez-Fernadez (2001). 
A detailed account of the main properties and modelling of \grss can be 
found in the review paper by Fender \& Belloni (2004).

Using a large collection of RXTE observations, Belloni et al. (2000) 
defined 12 different variability classes of X-ray emission, each of them
characterised by a time profile and spectral variability inferred from
the dynamical hardness ratio plots.
This classification is potentially useful for describing the behaviour of
this exceptional source and for understanding the physical processes
responsible for the X-ray emission, so we will refer to it when presenting 
our results.
These 12 variability classes, however, do not exhaust the very rich set of
temporal patterns exhibited by \grs: Klein-Wolt et al. (2002) and Hannikainen
et al. (2003, 2005) reported two other variability classes.
Furthermore, \grss exhibits a wide variety of complex behaviours when
class transitions occur.

\grss was observed by the Narrow Field Instruments (NFIs) onboard
the \sax~ satellite (Boella et al. 1997a) on several occasions
(Feroci et al. 1999, 2001; Ueda et al. 2002). The
amount of data collected during these pointings is very large and
requires a long and complex analysis.
In the present article, we report the
results of an analysis of the time behaviour of \grss observed in the
course of a long pointing in October 2000. On that occasion the
source was mainly observed in the $\rho$ class (Belloni et al.
2000), which is characterised by a series of bursts, with a variable
recurrence time from 40 to more than 100 seconds. The time
profile of the bursts has a rather smooth rising branch (also
named a `shoulder') followed by a series of intense and short peaks
with a fast decline.

The first observation of \grss in the $\rho$ class reported in the
literature was performed on 15 0ctober 1996 with Rossi-XTE  (Taam,
Chen \& Swank 1997) and the bursts exhibit a typical structure with two and
more peaks. Their typical profile in the 2--13 keV
band also differed from that in the 13-60 keV band. Belloni
et al. (1997a,b) and Taam, Chen \& Swank (1997) interpreted this
bursting activity as the result of thermal/viscous instabilities
in an accretion disk, in general agreement with the expectations
of previous theoretical calculations (Taam \& Lin 1984).

Other RXTE observations of the same class were described by Vilhu
\& Nevalainen (1998), Paul et al. (1998), and Yadav et al. (1999),
who presented the results of a long pointing of \grss performed in
June 1997 with the IXAE onboard the IRS-P3 satellite. On that
occasion the source was in the $\rho$ class for about five days,
then changed to the $\kappa$ class in which it remained for a
similar time and afterwards returned to the former one. Transitions
from/to the $\rho$ class to/from other classes were described
by Naik et al. (2002), Chakrabarti et al. (2004, 2005), and
Rodriguez et al. (2008).

A complete analysis of the timing and spectral properties of the
$\rho$ class has yet to be performed. The long and nearly continuous
data set, not yet published, presented and analysed in
this paper is unique for investigating the behaviour of \grss in
this particular variability class.
Because of the richness of the data set, we applied several
techniques to obtain an as complete as possible picture of the burst 
phenomenon, practically impossible to describe in a single
paper. In this work, we studied the timing properties
covering a range of timescales from a few seconds to a few days.
We first investigated the variations in the mean brightness in
different spectral bands on the timescale of a day and we then applied
Fourier periodograms and wavelet transforms to study the temporal
properties of the X-ray light curves. Finally, we investigated the
profile of the bursts, over a timescale of a few seconds, using a
statistical approach. We searched for correlations between
the timing behaviour of \grss and the X-ray brightness of the
bursts, finding highly significant relations. All of these results
are particularly useful for addressing either the spectral analysis or
studying the onset of unstable phases in terms of chaotic
processes. These subjects will be described in detail in a couple of
forthcoming papers.

\begin{figure}
\epsfxsize=9.0cm \epsfbox{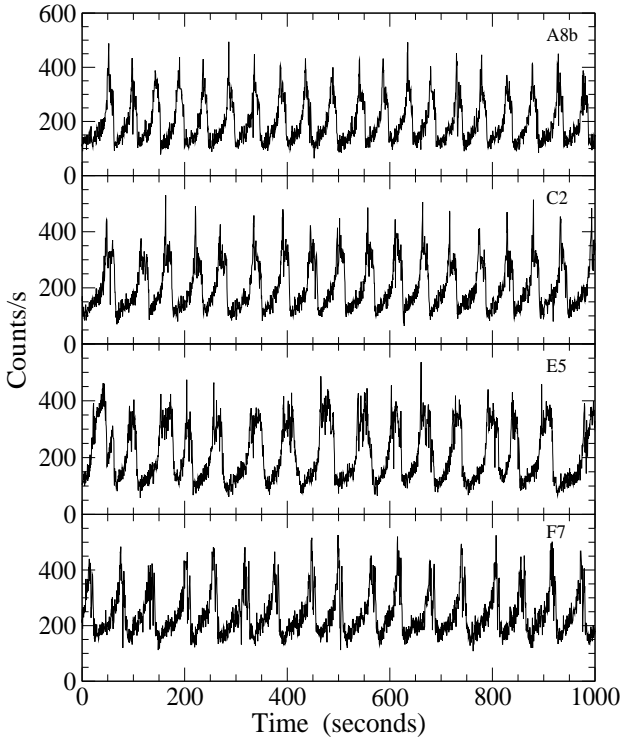} \caption[]{ Segments of
MECS (1.6--10 keV) light curves extracted from four data series
(top to bottom: A8b, C2, E5 and F7) at different times during the
\sax~ observations of October 2000. Time bin width is 0.5 s and
the starting time of each segment is arbitrarily selected within
the series. Note that the first two and the fourth panels exhibit the
characteristic time profile of the $\rho$ class, while the third
panel has a more complex structure.
}
\end{figure}

\begin{figure}
\epsfxsize=9.0cm
\epsfbox{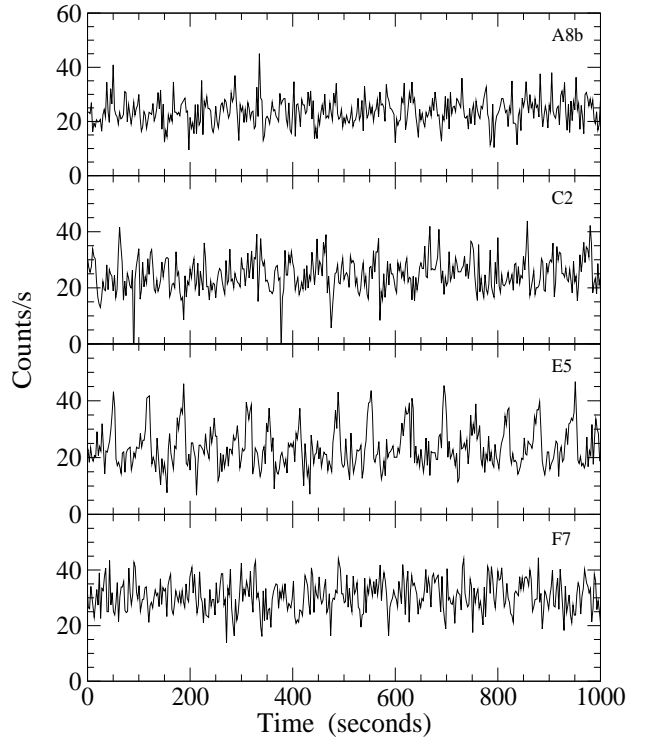}
\caption[]{
Segments of PDS (15--100 keV) light curves extracted from the same
times series of Fig. 1.
Time bin width is 2.5 s.
Note that individual bursts are more apparent in the E5 series (thid panel)
than in the other three.
}
\end{figure}

\section{The observations}

The \sax~ observation of \grss considered in the present paper
started on October 20, 2000 (MJD 51837.894) and terminated on October
29 after an overall duration of 768.79 ks.
The observation consisted of several runs of a typical duration of
about one day.
Each run is identified by a numeric code and consists of data
segments corresponding to the visibility intervals of the source along
the satellite orbit.
The data obtained with all the NFIs can be retrived from the \sax~
archive at the ASI Science Data Center.

In this paper, we limit our analysis to the data obtained in the first
six runs and partially in the seventh one for a total duration of
about 610,000 seconds, which covers about the 80\% of the entire
pointing. The net MECS exposure time is 257 ks, while the
PDS net exposure amounts to 123 ks. This choice is motivated by, 
in this period, \grss remaining mainly in the $\rho$
class with some phases of instability similar to the $\kappa$
class.
In the data acquared during the last few orbits of the seventh run and 
during the two remaining ones, the source changed to other and more complex 
classes and therefore, will be analysed in a forthcoming paper.
The log of the observation runs considered in the present
paper (see Table A.1) as well as the details of the data reduction are 
reported in Appendix I.

\begin{figure}
\epsfysize=11.0cm
\epsfbox{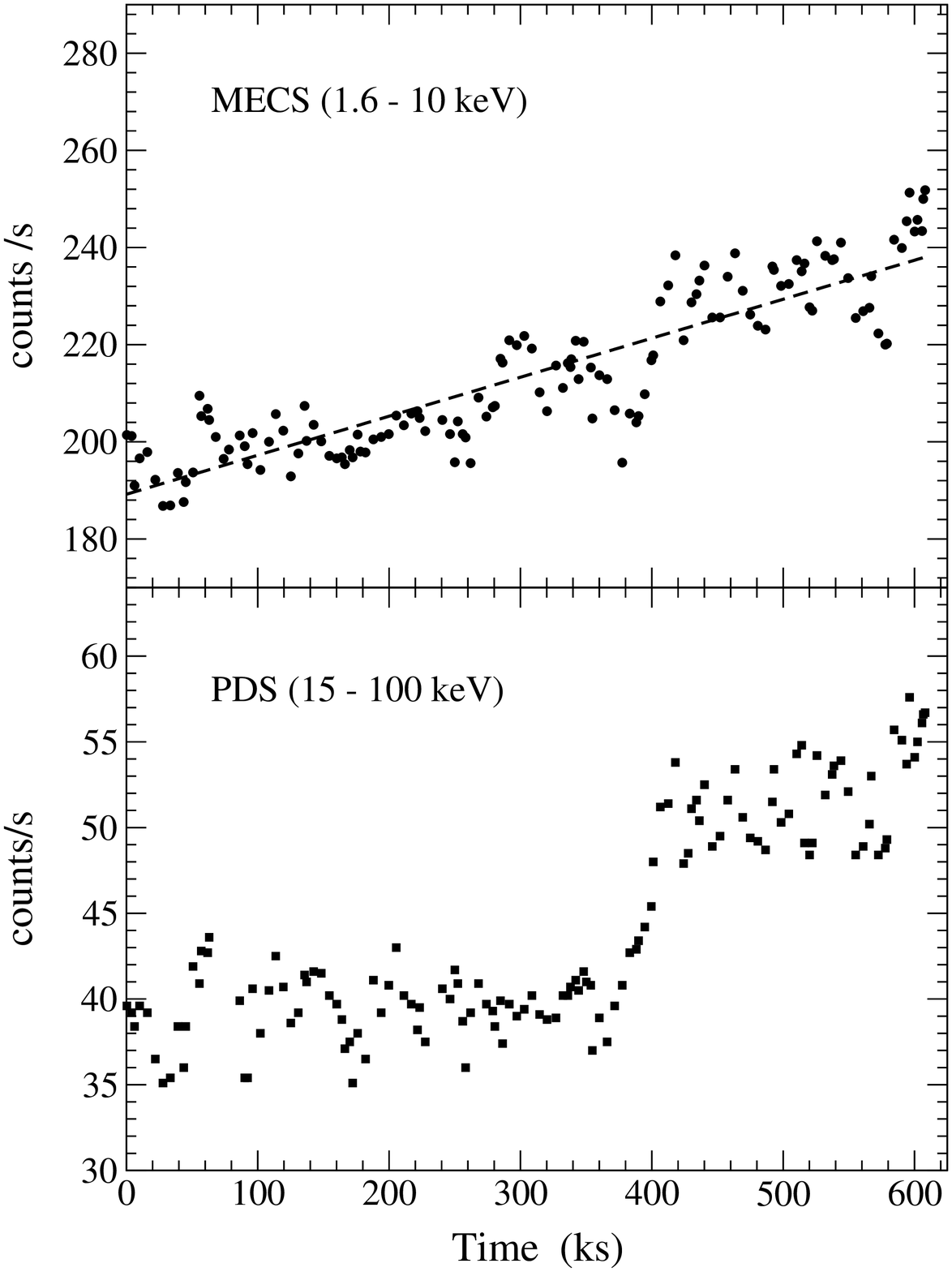}
\caption[]{
Long-term evolution of the X-ray count rates of \grss during the long observation
of October 2000 in the MECS (upper panel) and PDS (lower panel) energy bands.
The dashed line indicates the linear best-fit to MECS data.
In the PDS data, two brightness states are evident with a short transition between
them, occurring between 375 and about 400 ks.
}
\end{figure}

\begin{figure}
\epsfysize=11.0cm
\epsfbox{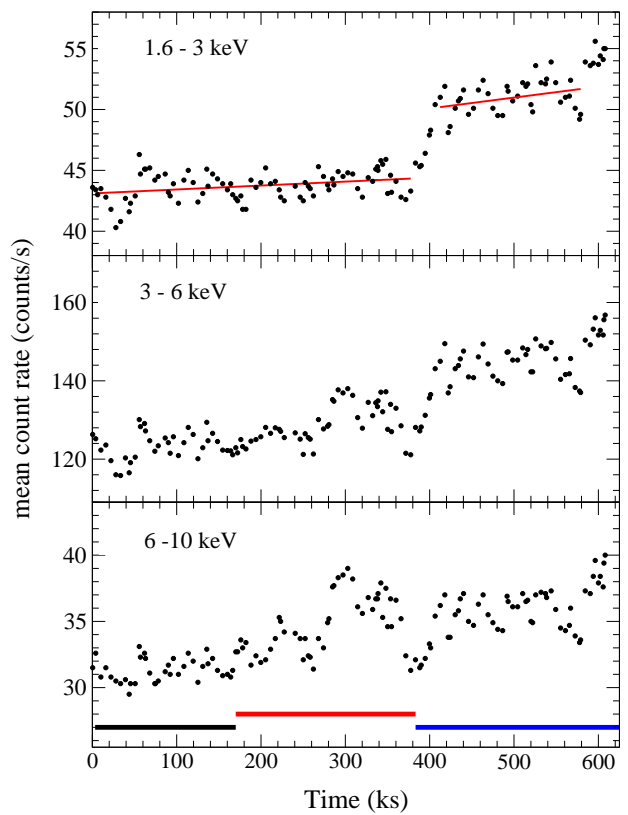}
\caption[]{
The time evolution of the MECS mean count rate in three energy bands:
1.6--3 keV, 3--6 keV and 6--10 keV (top to bottom).
The two lines (red in the electronic version) in the upper panel are the
linear best fits in the corresponding intervals.
The thick horizontal lines (black, red and blue in the e-version) in the
bottom panel indicate the three time segments used in our analysis of the
data: the second interval includes the local excess of count rate.
}
\end{figure}

\begin{figure}
\includegraphics[height=10.0cm,angle=0,scale=1.15]{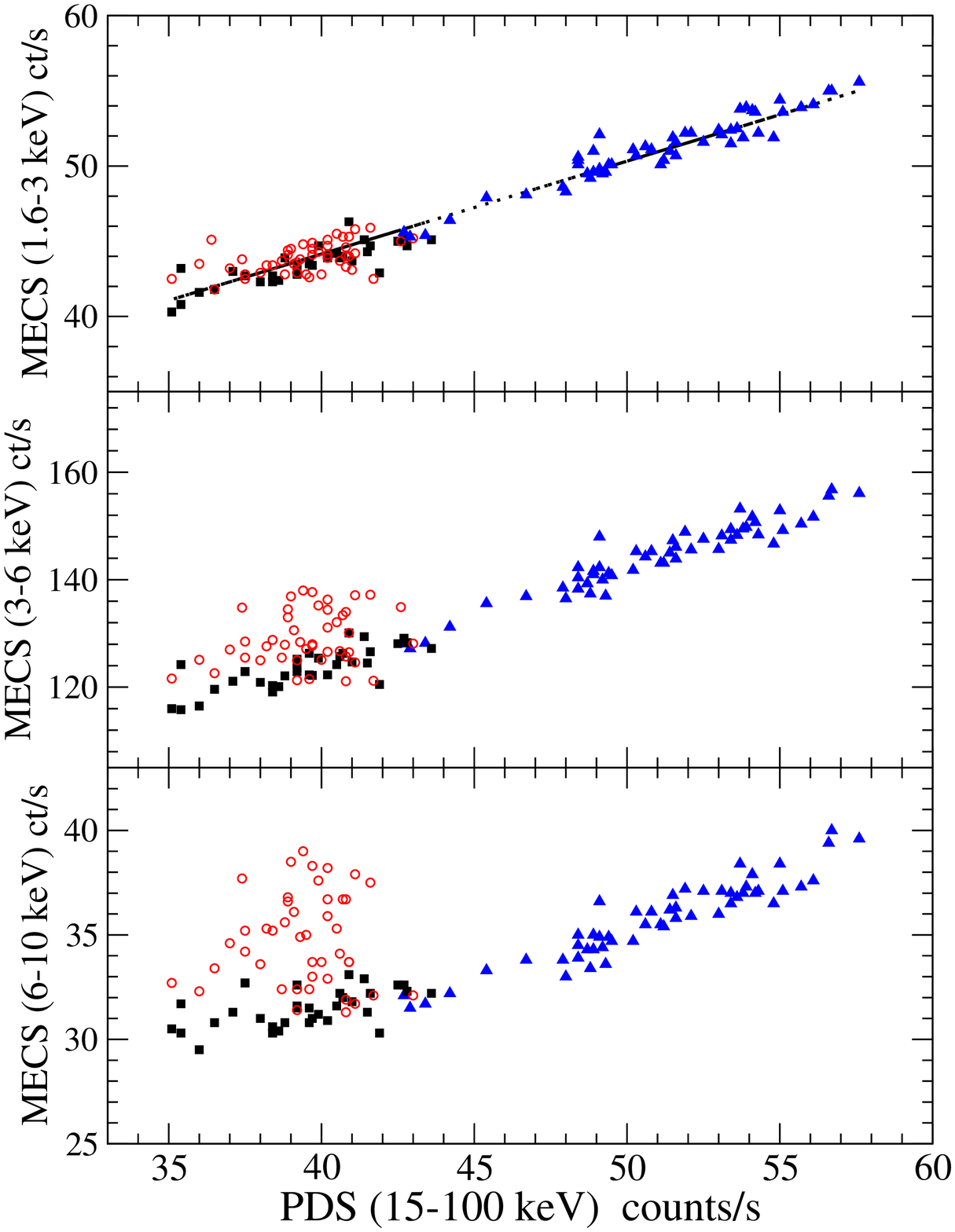}
\caption[]{ The mean MECS count rates in the three bands 1.6-3 keV
(top panel), 3-6 keV (middle panel), and 6-10 keV (bottom panel)
plotted against that of PDS. The three symbols and colours
corresponds to the data of the three time intervals of the MECS:
first interval (black filled squares), second interval (red open
circles), third interval (blue filled triangles). }
\end{figure}

It is practically impossible to describe completely this enormous 
amount of data and therefore we developed a rather simple classification 
scheme for the timing properties of individual segments based on Fourier 
and wavelet analyses.
To achieve a clearer understanding, however, we selected some data series
that are representative of the different behaviour of \grss and used them
as examples throughout this paper. These series are named A8b, C2,
E5, and F7 (see Appendix II and Table A.2) and some portions of them
are of equal length, 1,000 seconds, as shown in Fig. 1, for the
MECS and in Fig. 2 for the PDS, for which a time binning greater
than MECS was used to increase the signal to noise ratio (S/N). 
It is possible that the source exhibited the behaviour characteristic of the
$\rho$ class of Belloni et al. (2000). We note, however that the
light curve in the third panel (E5) exhibits behaviour
intermediate between this and the $\kappa$ class,
characterised by broad and slow bursts. The S/N of PDS data
is poorer than MECS and the bursts of A8b, C2, and F7 segments are
barely apparent, while those of the E5 segment are all clearly
distinguishable. As discussed later, this finding is a
consequence of a spectral evolution related to the behaviour of
the source; we also note that bursts in PDS series appear {sharper than
those of the MECS}.

In the various panels in Figs. 1 and 2, there is an important indication
of the change in the mean flux of the source.
Both of the last data series (F7) have count rates that are significantly higher
than the other three.
In particular, the series in Fig. 1 shows an increase in the minimum level
of about 30\% from $\sim$120 counts/s in the series A8b to $\sim$165 counts/s
in F7, while the bursts' height does not show a comparable increase.

\section {Slow variations}

The X-ray flux from \grss varied during the observation, exhibiting
a generally increasing trend with superimposed variations on
timescales of a few hours. Changes occurring on the
scale of one day or longer (slow variations) are evident
from the evolution in the count rate in the MECS and PDS bands
averaged over each time segment. Figure 3 shows the count rate
history of the MECS (1.6--10 keV) and PDS (15 -- 100 keV) energy
bands. The mean count rate of the MECS, computed only from time
series longer than 500 seconds, increased from $\sim$200 counts/s
at the beginning of the observation to reach $\sim$250 counts/s at
the end. It can be described by a linear interpolation (the linear
correlation coefficient is $r$=0.884) with a positive rate of
$\sim$7 (counts/s)/day.

The evolution in the mean PDS count rate appears to be different from that
of the MECS and shows two rather long states of different intensity: 
in the first one, the mean count rate varied between 35 and 44 counts/s,
whereas in the second state it was between 48 and 57 counts/s. These
two states are separated by a relatively fast transition that
started at about 375 ks after the beginning of the pointing and
lasted about 30 ks, thus including the time series from E7 to E12.
The mean PDS count rate changed from the average value 39.4 and a
standard deviation of 1.9, to 51.7 and 2.7, respectively, with
a corresponding luminosity increase above 15 keV by about 30\%.

To obtain a clearer description of these different behaviours, it is useful
to consider the count rate evolution in three narrow bands of the
MECS, of nominal energies of 1.6--3, 3--6, and 6--10 keV
(the correspondence with the true photon energy being good because
the instrumental response matrix is essentially diagonal) and
selected to have a sufficiently high S/N. The resulting
plots are shown in the three panels of Fig. 4 and some interesting
differences appear between them. The low energy plot is remarkably
similar to that of the PDS: the mean count rate was almost
constant around 43 counts/s up to $\sim$375 ks, then increased
over about 30 ks to $\sim$51 counts/s and fluctuated
around this high level until the end. In the intermediate and high
energy ranges, the count rate evolution is much closer to the
general MECS trend. At energies higher than 6 keV, however, the
count rate in the central part of the observation had a mean level of
around 38 counts/s, higher than in the previous portion ($\sim$32
counts/s). This high level finished just before the fast increase
in the low energy and PDS curves. A similar change can also be
recognized in the intermediate band, but is absent both at
lower energies and in the PDS. To take this feature into account, 
we divided the entire observation into three intervals: the first
interval extends from the beginning to about 170 ks and includes
the series up to B15a, the second from $\sim$170 ks to $\sim$380 ks,
includes the entire bump in the 6-10 keV band (series B15b to E7),
the third interval covers the remaining time from the series E8 until
the end.
These intervals are indicated by the three horizontal
lines in the bottom panel of Fig. 4.

The differences between the various energy bands become clearer in
plots that compare the mean count rates in three MECS bands with
in the PDS band (Fig. 5).
As expected from the similar time evolution, the 1.6--3 keV
MECS and PDS count rates have a strong positive correlation: the
linear correlation coefficient is $r=0.974$, confirming
the very tight relationship. Similar trends are also exhibited by the
count rates in the two other MECS bands with the exception
of the data points in the second interval, which do not follow the
correlation and show an increase in scatter with energy. There
are, however, some points in the second interval with a rather low
count rate that are mixed with those of the first interval.
These series occurred in the first part of the interval and their
behaviour was remarkably similar to that which is typical of series in the
first interval. These results can be assumed to be indicative of a close
relation between the processes responsible for the emission below $\sim$3 
keV and above 15 keV, while in the range (3--15 keV) an
additional contribution of a different origin can occasionally be
observed.

\section{Fourier and wavelet classification}
The first approach to investigating the timescales of the burst
sequence in each data series is to derive the Fourier power
spectra or periodograms. The different appearance of these
periodograms lead us to introduce a practical classification of
series that is useful to a synthetic description of the variability
in \grss in the course of the observation. This classification is
also useful to the study of the chaotic states that appear in the
complex limit-cycle phenomenology. We argue that the \grss system
in the $\rho$ state can also exhibit states with chaotic
properties (see, the analysis of Misra et al. 2006), as we will
discuss in future work.

In several data series, however, there are large variations from
one burst to the subsequent one that are not described well by
the corresponding periodogram. To obtain more information about
variations occurring on scales of between a few tens and hundreds of
seconds, we applied the wavelet transform and devised a more
complete two parameter classification.

\subsection{Fourier periodograms and classification}

A large fraction of the Fourier periodograms (hereafter FP) exhibit
a single dominant peak, indicating that bursts occurred with a
rather stable {\it recurrence time}, \trec. In contrast other spectra
have two or more dominant features that do not correspond to the
harmonics of the main peak. We prefer to use the term recurrence
time instead of `period' because the burst sequence has never been
observed to remain very stable for time intervals of about one
hour, as shown by the wavelet analysis presented in Sect.
4.2.
In Table A.3, we reported the values of \trec and
the corresponding FP time resolution $\Delta t$ that is an
estimate of the uncertainty. As illustrated by the classification
scheme presented in Appendix III, we introduced three types of periodograms
according to the power distribution: S type when only one dominant
peak is apparent in the FP, T type when two peaks are present, and
M type for a higher peak number. Figure 6 shows the FPs of the four
MECS data series plotted in Fig. 1: the first FP is of S type, the
third of M type, and the remainder are both of T type with
different peak heights and separation.

\begin{figure}
\epsfysize=12.0cm
\epsfbox{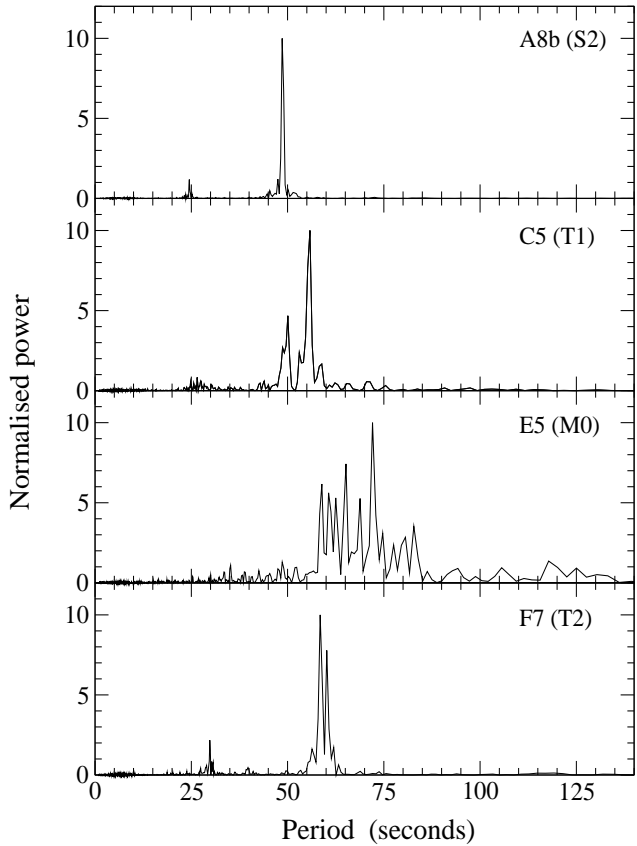}
\caption[]{
Periodograms of four MECS time segments shown in Fig. 1.
The spectral classification based on Fourier periodograms and wavelet
scalograms in reported in parenthesis.
}
\end{figure}

\begin{figure}
\epsfysize=12.0cm
\epsfbox{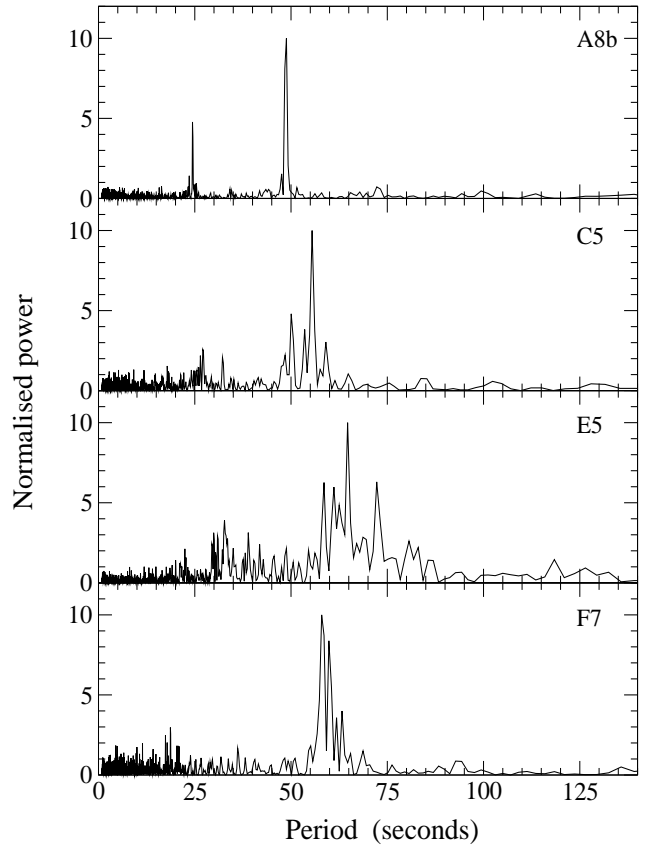}
\caption[]{
Periodograms of four PDS time segments (from top to bottom) A8b, C2, E5, F7.
Note the similarity of power distributions with those of Fig. 3 despite the
lower S/N.
}
\end{figure}

We computed the FPs for the PDS time series with durations
comparable to those of the corresponding MECS series.
Their classification, however, is not as good as for the MECS because of
the low S/N ratio.
Figure 7 shows the FPs of the already considered four series: significant
peaks are present at the same periods as those of the MECS series,
although the power distributions between the peaks differ.
We note that the power at the first harmonic in the PDS periodograms is
generally higher than in the MECS.

An interesting result is the change of \trec~ in the course of the
pointing. In Fig. 8, the values of \trec~ are plotted as a function
of time for the three intervals using different symbol in order to
distinguish the type of each series. 
In the first interval,
\trec~ is rather stable, while in the second the dispersion of
\trec~ and the width of the peak ranges of M series are much
higher than in the other two.  We note that, as in Fig. 5, some series
in the initial portion of this latter interval, have \trec~ values
close to those found in the first interval. In the third interval,
the recurrence time exhibits a generally increasing trend.

As shown in Sect. 3, the mean count rate increases during the pointing
and one can therefore expect that it should be correlated with \trec.
In Fig. 9, \trec~ is plotted against the MECS (for the two bands 1.6--3
and 3--10 keV) and PDS count rates.
In the second interval, no correlation is apparent, while it is clearly apparent
in the third interval.
The linear correlation coefficient is around $r=0.89$  for all three
energy bands. 
We note also the similarity between the first and the third panel of Fig. 9,
which confirms the strong relation between the lowest MECS
and PDS energy ranges.

The FP classification does not provide a univocal description
of the source behaviour: in a few cases, we found series with
an apparently irregular sequence of bursts (e.g., D3a or D8b) but the power
mostly concentrated in a single peak.
It was therefore necessary to improve the classification by taking 
account of how the power is distributed over the various timescales in
the course of the series, whereas the standard Fourier analysis considers
the entire series.
This analysis can be performed by means of wavelet transforms and the results
are described in the following.

\begin{figure}
\includegraphics[height=9.0cm,angle=-90,scale=1.0]{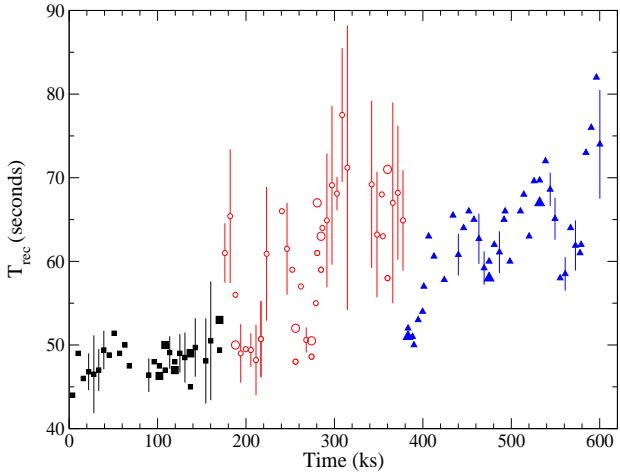}
\caption[]{
The evolution in the recurrence time of the MECS series in the course of the
observation.
Along the abscissa, we plot is the time elapsed since the beginning of the observation.
Symbols indicate series of three time intervals: first interval (filled black
squares) including series from A1b to C6, second interval (open red circles)
including series from C7 to E7, and a third interval (filled blue triangles)
including series to G12c.
For the T series, the recurrence times of both peaks are plotted (using 
symbols of different size), whereas the points of M series correspond
to the centroid value as given in Table A.3 and the vertical bar to the amplitude
of the range.
}
\end{figure}

\begin{figure}
\epsfysize=12.0cm
\epsfbox{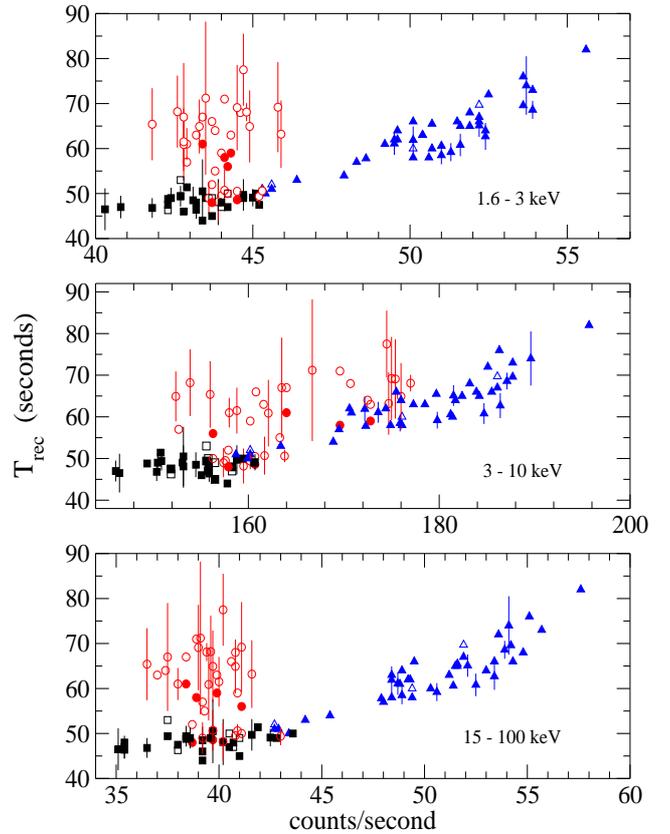}
\caption[]{
The relations between the recurrence time and the mean count rates
in the energy bands.
Top to bottom: MECS 1.6--3 keV, MECS 3--10 keV, PDS 15--100 keV.
Data symbols correspond to the three time intervals defined in
text (see also Fig. 4).
}
\end{figure}

\subsection{Wavelet spectra and their classification}

The wavelet transform permits us to decompose a signal using a
localized function. Standard wavelet analysis is based on the
computation of wavelet power spectra (also named {\it wavelet
scalograms}, hereafter WS) defined as the normalised square of the
modulus of the wavelet transform. A short description of the
algorithm and definitions of the spectral quantities are given
in Appendix IV. An advantage of this local analysis is that
scalograms are less sensitive to telemetry gaps than FP and it is
possible to consider longer time series.

The WSs for three of the MECS time series in Fig. 1 (A8b, E5, F7)
are presented in the left-hand side panels of Fig. 10:
each spectrum consists of a two dimensional map where the curves of equal
power define regions coded in color/grey scale.
Because of the smoothing and interpolation necessary to draw these curves, 
these plots provide a qualitative description of the evolution of the
power during the time series.

The differences between both the S and T series and the M series are
clearly apparent: the two former types are characterised by an uninterrupted
and nearly horizontal strip, centred on a timescale length close to the
\trec~ estimated from periodograms.
The coarser resolution of WSs compared to FPs does not allow us to
identify clearly the peak separation in the T series, which is only 2 s
(panel 3).
However, a decrease in the highest power timescale is discernible 
around 1700 seconds.
Another strip of lower intensity is centred around the half value
of \trec~ and corresponds to the first harmonic.
Small changes in \trec~ are also present in the WS of the S series,
although their duration is limited to only a few bursts, too short
to produce a separate peak in the FP.
It is unclear whether there is a sort of slow modulation of the
recurrence time of subsequent bursts. In some cases, these
changes occur on timescales longer than the duration of
two or three bursts.

\begin{figure*}
      \hspace{3cm}
\includegraphics[height=10.0cm,angle=0,scale=0.90]{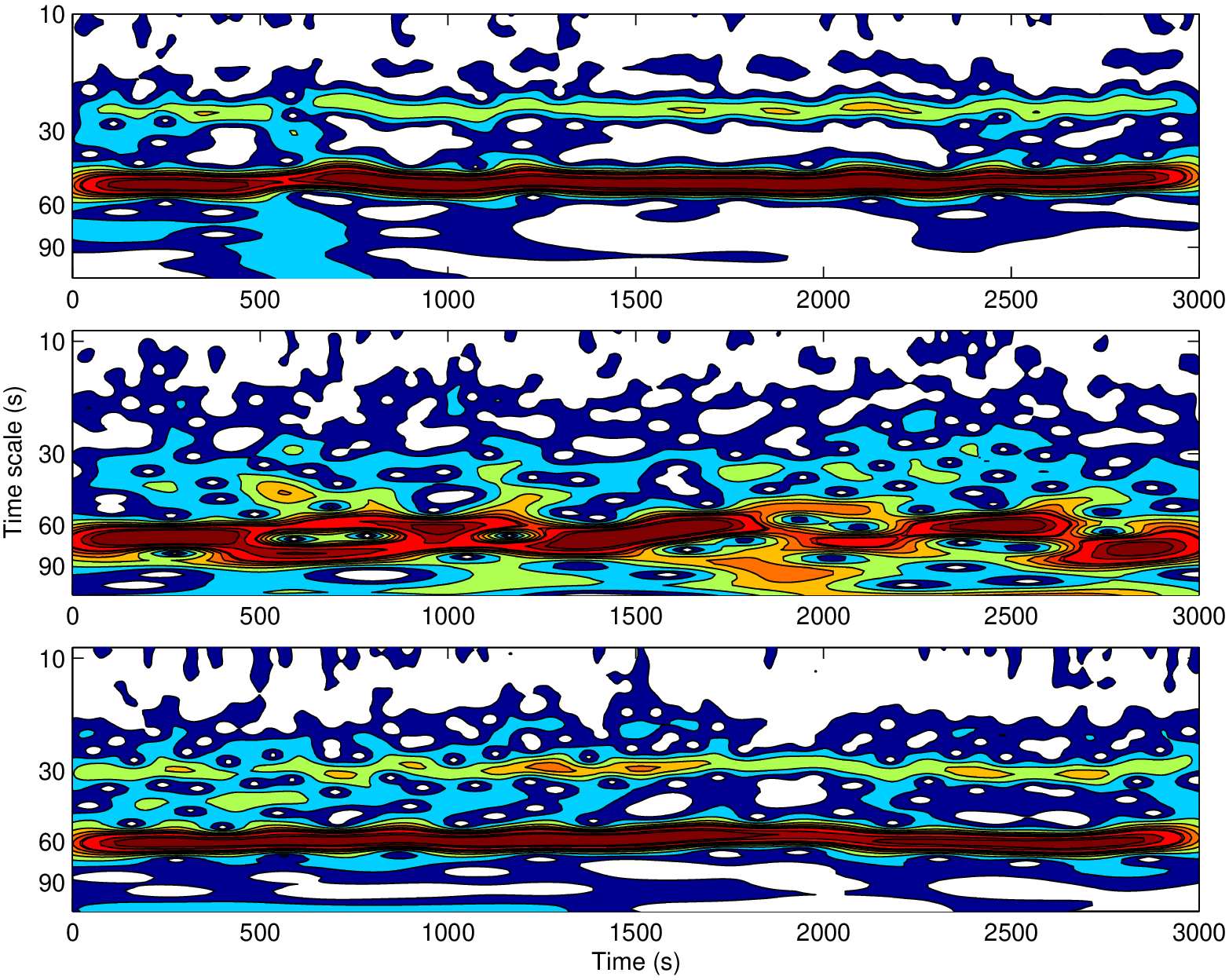}\includegraphics[height=10.0cm,angle=0,scale=0.90]{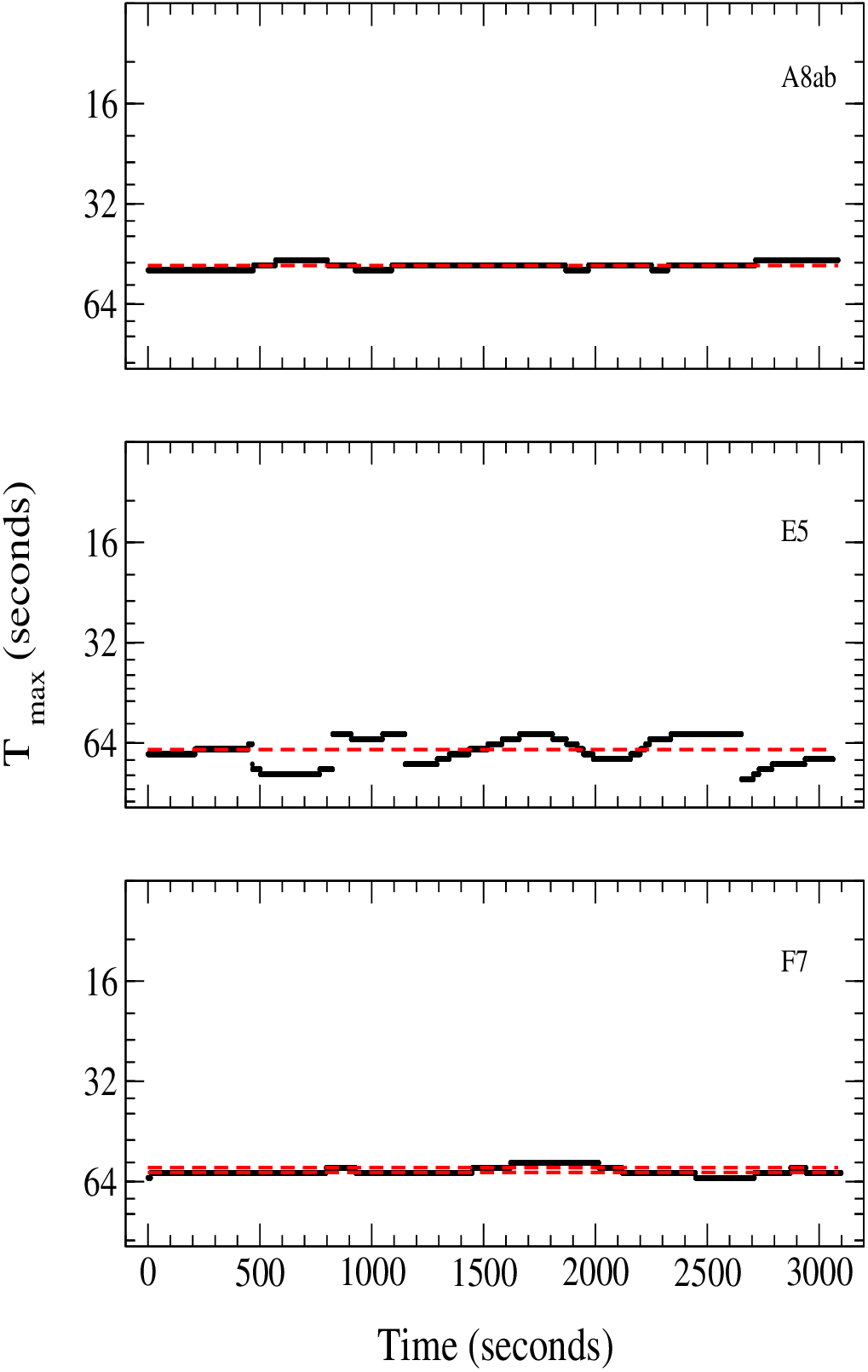}
\caption[]{
Left panel: iso-power representation of the scalograms of three MECS time series
A8ab (the two series are joined together), E5 and F7 (top to bottom), plotted
in Fig.1.
In each spectrum, the time scale increases towards the bottom.
The dark strip indicates the region where the highest power is found and
corresponds to \trec; the strip of the first harmonics is also
clearly visible in the top and bottom panels, whereas it is
not present in the highly irregular E5 series.

Right panel: time evolution of the timescales of the power maxima in the
scalograms of the three series in the left panel:
$T_{j,max}$ increases with a log scale towards the bottom to permit a 
clearer comparison with the spectra in the left panel.
Dashed lines represent the values of the \trec~ peaks or
the peak centroid from Fourier periodograms.}
\end{figure*}

The M series have typically a far more irregular pattern also over
short time intervals.
As for the E5 series in Fig. 10, power is concentrated within a rather
broad strip that exhibits meanderings, bifurcations, and interruptions
corresponding to short and long recurrence intervals between subsequent bursts.
A fraction of the power that is greater than in the two other cases is present
on both long and short timescales, and no strip corresponds to the first harmonics.

To evaluate the stability of the time series from WSs, we adopted a
criterium based on the relative variance ratio $R_w$ (see Appendix
IV). There is a correspondence between the values of $R_w$ and the
periodogram classification: this parameter tends to increase between
the series S and T, and between the series T and M.  %%% diverso da LE 
The presence of a single dominant
peak is not necessarily indicative of a stable signal: a series
can exhibit large instabilities occurring within a rather small time
interval or frequent changes in \trec~ but with a
rather stable mean value. On the other hand, more peaks in the FP
occurring with a rather narrow range can correspond to small
variations in the recurrence time without a large modification of
the bursts' structure. To take this variety of behaviour into account 
we defined three classes, indicated by 0, 1, and 2 in order of
increasing stability (see Appendix IV).
The majority of S spectra, 30 among 45, are of S2 type, 12 are S1, and 3
S0; we also note that 10 of the S1 spectra have $R_w$ smaller than 4.0,
indicating that their \trec~ is only moderately unstable.
In contrast, M series are of class 0 and 1, and only four are of class 2
but with $R_w$ values larger than 2.5.
Nine spectra of T series are of class 1, confirming their intermediate type 
between S and M.
The complete classification of the spectra in Fig. 10 is indicated by the labels
of Fig. 6.

The results presented in this and previous Sections suggest that
one can consider two modes of the $\rho$ class, which we call {\it regular} and
{\it irregular} modes.
The former is associated with the occurrence of S or T spectra of stability
class 1 or 2, the latter with M series or stability class 0.

\section{Analysis of the burst structure}

It is also important to study how the burst structure
changes in the various modes.
As pointed out by Belloni et al. (2000), the individual bursts, when analysed
with sufficient time resolution, exhibit several substructures, such as 
sharp spikes and dips, of durations as short as to a few hundreds of milliseconds, and
probably even shorter.
These substructures are highly variable and therefore it is practically
impossible to study the structure of thousands of bursts with a high time
resolution.
We adopted the two following approaches:
we first performed a statistical study of a large sample of bursts to describe
their mean properties, and afterwards studied the structure of some typical
bursts of two of the selected time series to take account of
the energy dependence.

\begin{figure}
\includegraphics[height=9.0cm,angle=-90,scale=1.37]{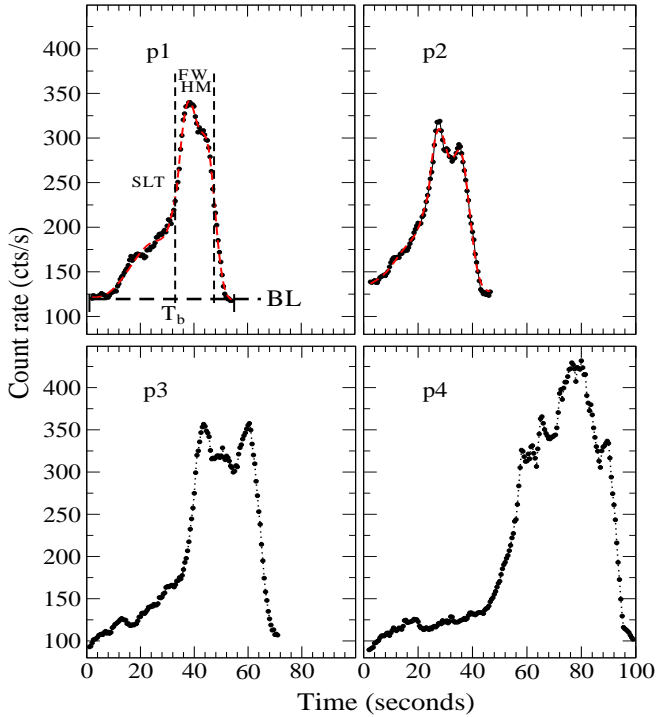}
\caption[]{
Four examples of burst shapes after smoothing with
peak number classification from {\it p1} to {\it p4}.
Best fits using 2 Gaussian components for the peak and a fourth
degree polynomial for the slow leading trail are shown for the two upper
bursts (dashed/red line).
}
\end{figure}

\begin{figure}
\includegraphics[width=8.5cm,angle=-90,scale=1.37]{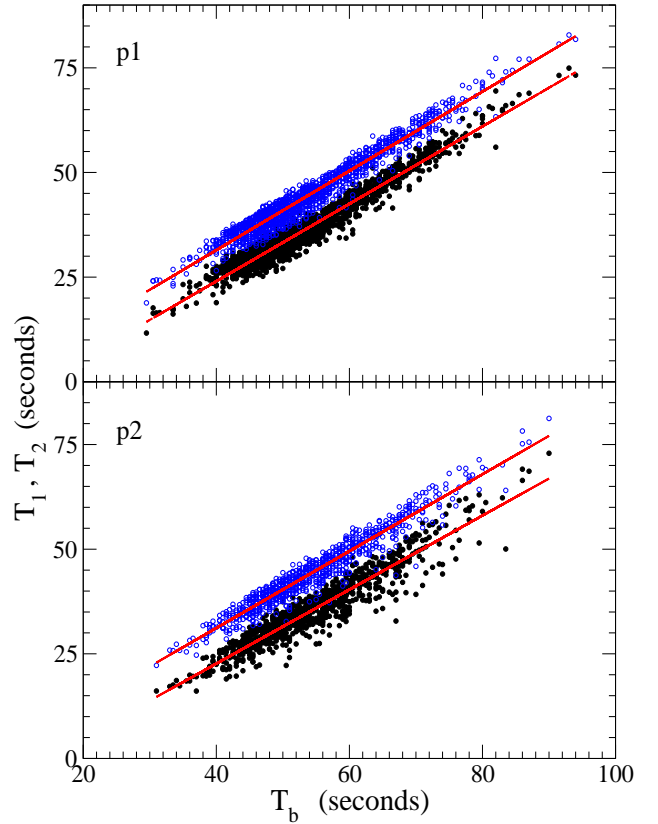}
\caption[]{ Plot of $T_1$ (black solid circles) and $T_2$ (open
blue circles) versus $T_b$ showing the strong linear correlations of
the central times of the two Gaussian curves with the total
duration of the bursts. Times for {\it p1} pulses are plotted in
the upper panel and those for the {\it p2} pulses in the lower
panel. Linear best fits (in red) are also shown for each data set.
The noticeable similarity of these data distributions indicates
that there are no significant differences between these
multiplicity classes. }
\end{figure}

\subsection{The statistical analysis}

To avoid the complications caused by the rich and variable substructures,
a statistical analysis was performed after a running-average
smoothing of all data series for the entire MECS energy band with a
window of 5.5 seconds (11 time bins). Despite the complex
structure, the burst shape after smoothing appears rather
stable. We introduced peak number or ``multiplicity'' classes
based on the number of distinguishable features in the smoothed
profile. Examples of bursts of different multiplicity from 1
(hereafter indicated as {\it p1}) to 4 ({\it p4}) are shown in the
various panels of Fig. 11. This classification was applied to all the 172
time series, for a total number of 4083 bursts, also because 
the series that are too short for the Fourier analysis still contain a
large number of useful bursts.

We measured the duration $T_b$ of individual bursts starting from
the initial minimum level, which is reached just after the decay tail of
the preceding one. This minimum level was verified to be nicely
constant in the course of each series. It defines a \emph{baseline
level} (BL) over which the bursts are superimposed.
We also divided their typical structure into two parts: the shoulder
or \emph{slow leading trail} (SLT), which is followed
by a \emph{Pulse}
(see examples of this partition in Fig. 11).
For a small number of bursts, the identification of the end was 
tricky because there is a final and clearly separated pulse that occurs 
before the count rate has reached its lowest level. We called these bursts {\it
anomalous} and excluded them from the shape analysis. A couple of
anomalous bursts are clearly evident in the light curve of the E5 series
shown in Fig. 1. Anomalous bursts are observed in the {\it
irregular} mode and very rare or absent in the {\it regular} one.

\begin{figure}
\includegraphics[height=7.0cm,angle=-90,scale=1.37]{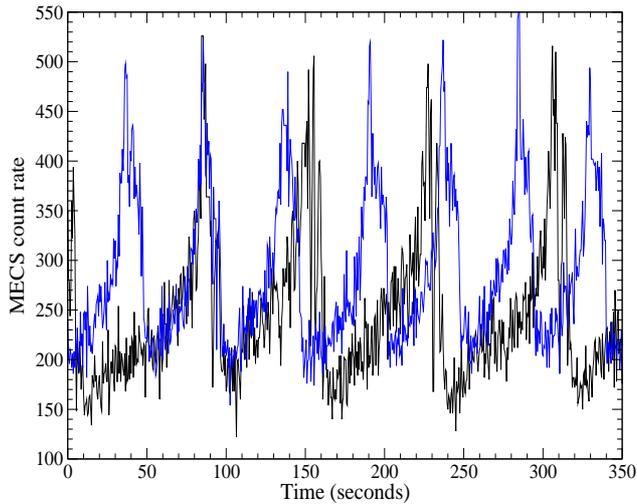}
\caption[]{
Comparison between two short segments of equal duration of the S2 series
A8b (blue) and G9 (black) time series characterised by very different \trec~
and mean count rates.
The A8b light curve has been shifted upward by 90 ct/s to match the peaks'
height.
Note the very good agreement between the pulse shapes of the two series,
in contrast to the leading edge being much longer in G9 than in A8b.
The bin width is 0.5 s.
}
\end{figure}

\emph{Pulses}, examined at high time resolution, have  typical structures
consisting of a sequence of peaks of variable height and duration, but when studied
over a timescale of a few seconds they have more stable profiles, such as
those in Fig. 11.
Pulses of {\it p1} and {\it p2} bursts were modelled by a combination of two
Gaussian curves and a fourth degree polynomial was used to reproduce SLT,
which terminated at the half maximum height of the pulse.
Best-fit curves are shown in the upper panels of Fig. 11: the need of the
two Gaussian components for both {\it p1} and {\it p2} bursts is evident.
We did not calculate best fits for bursts of type {\it p3} or higher because
they required additional Gaussian components and in some cases we did not obtain
stable solutions.

A first relevant result of this analysis is that the central times $T_1$
and $T_2$ ($T_2 > T_1$) of the two Gaussian components, measured starting
from the initial time of the burst, are strongly correlated with the
duration of individual bursts, measured to be the time
separation $T_b$ between the minima of two subsequent bursts.
The average $T_b$ in regular time series is very close to \trec.
Figure 12 shows a scatter plot of $T_1$ and $T_2$ against $T_b$ for
the two considered multiplicities:
%{\it p1} and {\it p2} bursts:
the points are very well aligned along straight lines
and no significant difference is evident between the two multiplicity
classes.
Linear correlation coefficients are very high ranging from 0.944 to
0.973.
Best-fit linear relations have very similar slopes (within $\sim$3\%) 
and this means that the difference $T_2 - T_1$, which can be considered to be
an estimate of the pulse width, is practically independent of $T_b$.

\begin{figure}
\includegraphics[height=9.0cm,angle=-90,scale=1.37]{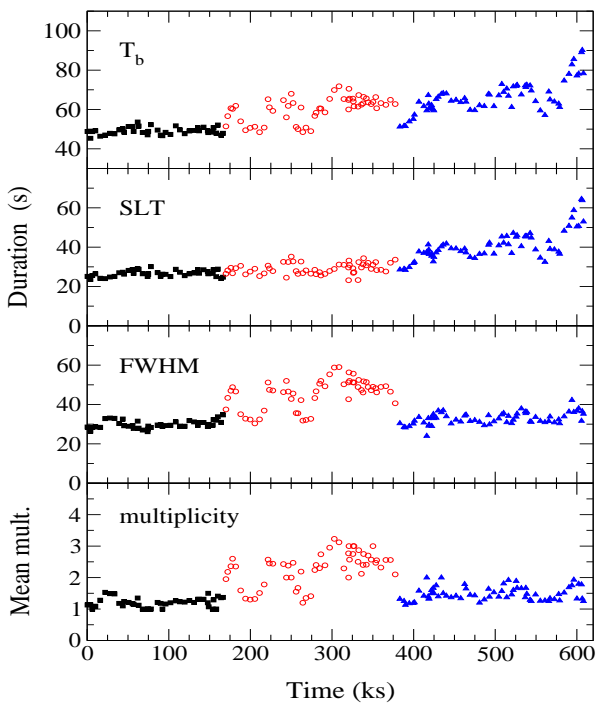}
\caption[]{ Time history of the durations of the various
components of bursts (top to bottom): $\langle T_b \rangle$, mean
SLT, mean pulses' FWHM. The scales on the ordinates are identical
to ease the comparison of data. In the fourth panel, we plot
the evolution in the mean multiplicity of bursts. Symbols that
identify the series in the three time intervals are the same
as those of Fig. 5.
}
\end{figure}

\begin{figure}
\includegraphics[height=9.0cm,angle=-90,scale=1.37]{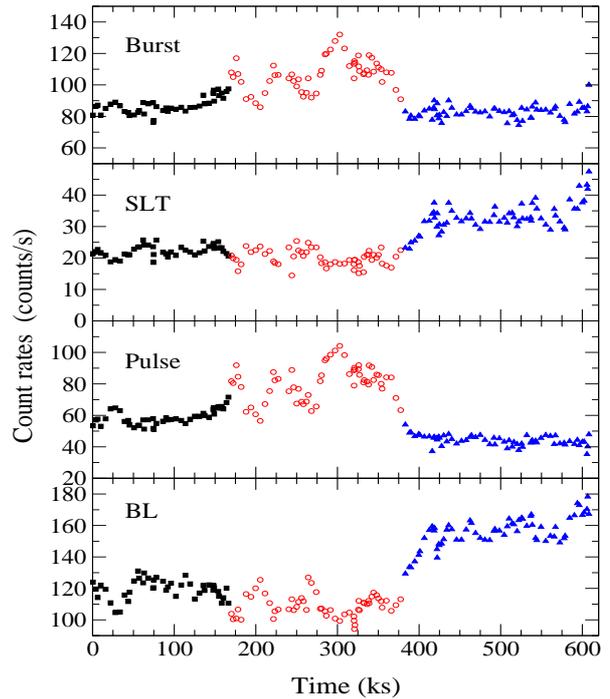}
\caption[]{ Time history of the mean MECS (1.6--10 keV) count
rates of the various components after the subtraction of the BL
(top to bottom): total burst, SLT, and the peak integral. The
behaviour of the BL is shown in the last panel. The count rate
scales in the three upper panels are equal, while that of the last one
has been doubled in length. Symbol codes indicate the three time 
intervals as in the other figures. }
\end{figure}

This result implies that series of type S2, but with mainly different
recurrence times, must have pulses of similar width.
To illustrate this statement, we plot in Fig. 13 two short segments of
the two S2 series A8b and G9:
\trec~ is 49 s for the former and 73 s for the latter.
Data were translated such that two bursts were superimposed two bursts: 
we note that they are of comparable height and width at variance with the SLT,
which is much longer in the G9 series than in A8b.

The evolution of the durations and count rates of the various
burst components, averaged over each data series, is presented in
Figs. 14 and 15, respectively. We assumed that only data series
with more than five bursts provide representative mean
values. For bursts of multiplicity {\it p3} or higher, for which
no best-fit modelling was obtained, the durations of the FWHM were  %%%%
estimated directly from the smoothed profiles. The first panel in
Fig. 14 shows the evolution of the mean $T_b$, which for the S and
T series is similar to that of \trec~ (see Fig. 8), while the
latter has a larger scatter for the M series. The duration of the
SLT (second panel in Fig. 14) was far more regular: it increased
very slowly from the beginning of the observation and this trend
did not show appreciable changes in the second interval, while a
moderate increase in the rate occurred in the third interval,
which became higher one only in the last 40 ks. The mean FWMH of
pulses (third panel), which was nearly constant in the first interval,
showed large variations in the second interval: it increased for the
first time after 170 ks from the beginning of the observation and
returned to the previous values after a few series; it then
increased again at about 250 ks and remained high for about 40 ks;
a third high state was finally in the last and longest part of
the second interval. In the third interval, the FWHM of the peaks
returned to low values, but with a dispersion somewhat higher than
in the first one, particularly approaching the end of the
pointing. We also see in Figs. 14 and 15 that when the source was
in the irregular mode the multiplicity of bursts increased and was
strongly correlated with the \emph{pulse} count rate. We recall that this
particular behaviour represents the basis of the three interval
segmentation introduced in Sect. 4.

We also analysed how the mean count rates in the entire MECS band of
these components changed in the course of the pointing.
To calculate the count rates of SLT and of pulses, it was first necessary
to evaluate that of the BL, which for each piece of data was derived by
averaging the count rates in the bins with the lowest level at the
beginning of bursts.

The time history of the MECS BL count rate is shown in the bottom
panel of Fig. 15: it is remarkably similar to that of the both PDS
(Fig. 3) and the lowest energy range of the MECS (Fig.
4), despite the highest photon contribution is at energies between
3 and 10 keV. In particular, the mean BL did not show any increase
in the second interval, when the behaviour of \grss was
irregular.
Thus, there is no indication of a positive
correlation between the BL and the bursts' multiplicity and duration.
The evolution of the SLT and {\it pulse} count rates  are given in the
central panels of the same figure. In agreement with the previous
results on the mean durations of these components, the SLT is
highly correlated with the BL, while it is practically unaffected
by the pulse amplitude and multiplicity. The mean pulse count rate
has a very close correspondence with the multiplicity, and
decreases from $\sim$55 to $\sim$45 counts/s from the first to the
third interval, suggesting that it could be anticorrelated with
the BL and SLT behaviour.

Our results do not disagree with those of Naik et al.
(2002), who reported that the ``burst strength'' decreases for
increasing \trec. These authors define the ``burst
strength'' as the ratio of the peak count rate to its lowest level.
According to this definition, an increase in the BL and a rather
stable pulse intensity would reduce the burst strength.

\subsection{Energy dependence of the bursts' structure}

All the results derived in the previous analyses indicate that the
burst structure must be energy dependent, and that the SLT and pulses
should have different energy spectra.
This property can be more accurately analysed by considering light curves
at different energies.
In Figs. 16 and 17, we plotted two short segments of series A8b and E5,
respectively.
The three panels in each figure correspond to the energy ranges considered
above.
Changes in the light curve structure are evident: at the lowest energies
the count rate has a nearly triangular (or approximately sinusoidal) modulation
clearly associated with the SLT and the bursts are located on the
declining side.
At higher energies, bursts become sharper and sharper, while the triangular
modulation is reduced.
In the irregular mode, the pulse sharpening is even more evident above 15 keV.
Figure 18 compares two simultaneous segments of the MECS and PDS {\bf of} the E5 series
containing six bursts.
As already found by Paul et al. (1998), bursts in both series occur at the
same times, but those in the PDS are systematically narrower than those at lower
energies.
This effect cannot be entirely caused by the poorer S/N ratio of PDS data,
as is clearly evident, for example, from the burst occurring at 180 s or that
at 350 s.
Moreover, PDS pulses appear generally at the end of the MECS bursts
and this finding is in agreement with the spectral hardening in the course
of the bursts.
Paul et al. (1998) modelled the spectrum by a thermal emission from an accretion
disk and described this phenomenon in terms of increasing temperature close to the
end of the bursts.

\begin{figure}
\includegraphics[height=7.0cm,angle=-90,scale=1.37]{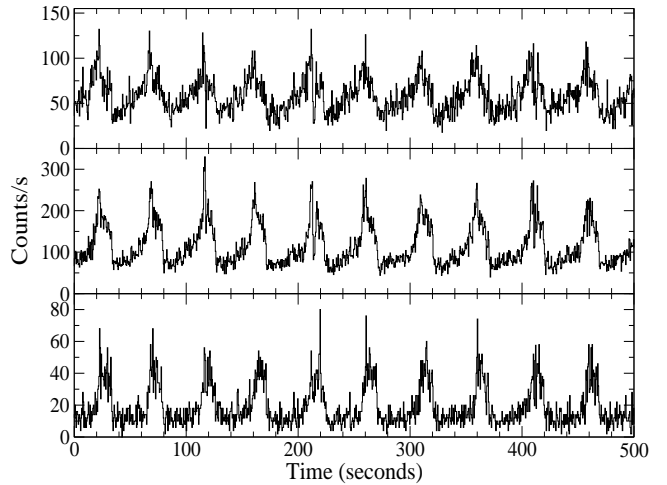}
\caption[]{
A short segment of the MECS A8b series in three energy bands: 1.6-3.0 keV
(upper panel), 3.0-6.0 keV (central panel), and 6.0-10.0 keV (lower panel).
The bin width 0.5 s for all sets.
Note that in the highest energy range the SLT has practically disappeared
and the pulses occur simultaneously with those at lower energies but their shape
is sharper and duration is shorter.
}
\end{figure}

\begin{figure}
\includegraphics[height=7.0cm,angle=-90,scale=1.37]{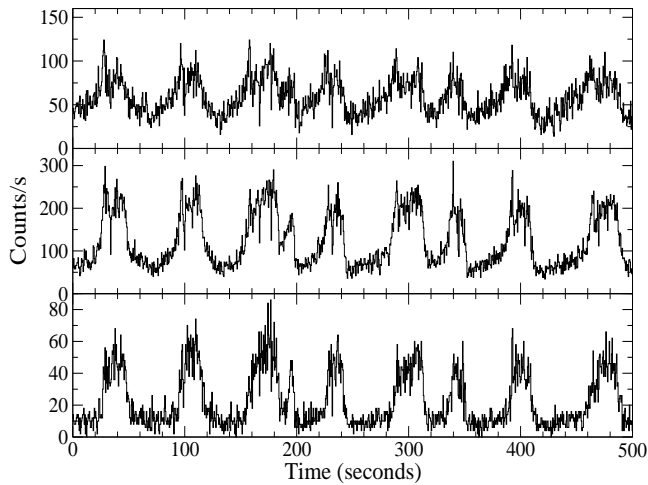}
\caption[]{
A short segment of the MECS E5 series in three energy bands: 1.6-3.0 keV
(upper panel), 3.0-6.0 keV (central panel), and 6.0-10.0 keV (lower panel).
The bin width 0.5 s for all sets.
Note that in the highest energy range the SLT hss practically disappeared
and the pulses occur simultaneously with those at lower energies but their shape
is sharper and duration shorter.
}
\end{figure}

To verify that bursts at higher energies are more sharply peaked than
at lower energies, we computed the autocorrelation functions of the
MECS and PDS E5 series because narrower bursts would correspond to
shorter de-correlation times (zero crossing). These found to be
equal to 16 s for the MECS and 10 s for the PDS. Similar
differences were also observed for other irregular PDS series, in
which individual bursts are clearly apparent.

\begin{figure}
\includegraphics[height=7.0cm,angle=-90,scale=1.37]{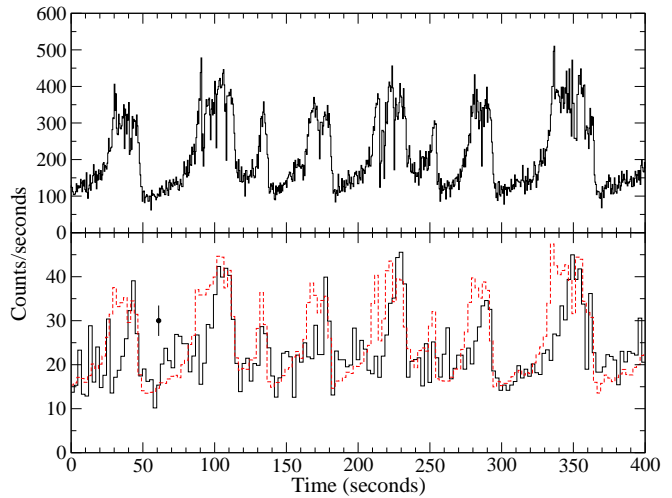}
\caption[]{
Comparison between MECS 1.6--10 keV (upper panel) and PDS 15--100 keV (lower panel)
burst shapes of the E5 series.
The latter has a bin width 2.5 s that reduces the statistical noise.
Dashed line in the lower panel is the same portion of the rebinned MECS light
curve scaled to PDS amplitude.
Note that bursts in the higher energy range occur simultaneously with those at
lower energies but their duration is generally shorter.
}
\end{figure}

\section{Summary and discussion}

In October 2000, \sax~ observed \grss continuously for about 10
days. For a large fraction of this time, the source exhibited the
characteristic behaviour of the $\rho$ class defined in Belloni et
al. (2000). In our analysis, the data obtained with the MECS and
PDS were organised into a database including many series of duration
corresponding to the spacecraft orbital windowing (or sometimes
shorter because of telemetry gaps). We analysed these time series
using Fourier periodograms and wavelet scalograms and introduced
some simple criteria, based on the number and height of prominent
peaks in the FPs and the variance ratio $R_w$ of scalograms, to
classify the behaviour of \grs. A complementary approach was the
study of the characteristics of individual bursts, based on the
definition of the two main components, namely the SLT (\emph{slow
leading trail}) and the \emph{pulse} superimposed on a stable BL.
Moreover, it was found to be useful to
introduce an additional multiplicity classification based on the
number of distinct features in the pulse.

We observed two different modes of the $\rho$ class: the
{\it regular} and {\it irregular} mode defined according to the following
characteristics.
The \textit{regular} mode scontains bursts with a nearly stable recurrence
time \trec~ and the pulses have a multiplicity of 1 or 2, and only
occasionally higher.
Pulses are clearly detected in the energy range between 3 and
10 keV and are weak at lower as well as at higher energies.
Below $\sim$3 keV, the shape of the signal modulation appears to be roughly
triangular, pulses are small and occur around the maxima.
In the \textit{irregular} mode, the burst sequence does not have a stable
recurrence time, the pulses have multiplicities generally greater
than 2, and anomalous bursts appear.
Pulses have harder spectra than in the regular mode, showing detectable
emission above 15 keV in the PDS data, but have typically shorter
durations than below 10 keV.
The irregular mode in several ways resembles the $\kappa$ class
(Belloni et al. 2000): it appears to be a transition state between
the regular $\rho$ and  the $\kappa$ class.

A classification of the burst types was introduced
by Yadav et al. (1999), who considered only some observations of
\grss that were typically shorter than one day. This work preceded the
definition of variability classes by Belloni et al. (2000), and the
light curves analysed by Yadav et al. (1999) belong to different
classes.

Differently from our work, they classified as irregular
the bursts of the $\kappa$ class, while as regular ones those of
the $\mu$ and $\rho$ classes.
They also found a correlation between the preceding quiescent time
and the burst duration for their irregular type, whereas
no such correlation was found for their regular ones. In
our analysis, we considered only the $\rho$ class, and this led us
to the discovery of some relevant relations between the typical 
timescales and the count rates in different energy bands.

The irregular mode was observed three times in our data (see Figs.
8, 14, and 15); during the rest of the observation, the source
maintained itself in a regular mode.
The first and shortest irregular phase started about 170 ks after
the beginning of the observation and lasted only $\sim$15 ks; the
second and third occurred between 220 and 260 ks and between 270 and
370 ks, respectively.
Interestingly, the last transition from irregular to regular mode
occurred in correspondence with a luminosity increase observed in
the MECS (1.6--3 keV) and PDS light curves (see Fig. 4).

The behaviour of the source throughout our observation exhibits
correlated signatures in both timing and intensity (i.e., count
rates) domains.
As an example, the recurrence time of bursts during the regular
phase was found to have a positive correlation with the average MECS
count rate (Fig. 9).
On the other hand, the source intensity in different energy ranges
exhibits a different behaviour: the rate below 3 keV correlates well with
that above 15 keV (Figs. 3 and 4), and both of them exhibits a transition
at the time $\sim$370 ks that also marks the last passage from the
irregular to the regular modes.
The intermediate energy range, 3--10 keV, appears to be unaffected by this
transition in the timing parameters.

We found that the BL and SLT exhibit remarkably similar behaviours and
the count rates are below 3 keV and above 15 keV. Instead, these
two components appear to be unaffected by the pulse multiplicity, which
is shown in our analysis to be capable of determining the duration of the 
{\it pulse} component. A signature of the time interval during which the 
multiplicity is highest (red open circles in the Figs. 14 and 15) is also 
clearly detected in the count rates of the \emph{pulse} component of the 
bursts (the higher the multiplicity the higher the \emph{pulse} 
count rate).

In the regular mode, we note that the increase in \trec~ is a
consequence of the SLT duration (see Figs. 13 and 14), while the
duration of the \emph{pulse} component remains nearly stable. This
implies that the characteristic timescales of these two components
follow different variability patterns.

This intermingled set of correlations can be interpreted by
introducing a two emission component scenario.
A `stable component' (shortly SC) is responsible for the
triangular/sinusoidal modulated emission observed below 3 keV, and
for the underlying BL emission (a clear separation between them is not
possible from the present analysis).
The other component, named the `pulsating component' (or PC) appears
with the pulses, which in the irregular mode they appear longer because of 
the higher number of peaks.
In addition, the peaks that appear last in a pulse have harder
spectra, because they are more evident above 15 keV than in the
regular mode.
It would be interesting to verify whether the onset of the PC is or is not
related to the level reached by the SC, but our analysis was 
inconclusive on this subject.

The count rate plots in the various bands indicate that the energy
spectra of SC and PC are expected to differ (see Figs.
16 and 17). In the 3--10 keV range, Taam, Chen \& Swank (1997) already 
noticed a significant difference between the spectra of
SLT and those of the pulses, the latter being harder, although in
their analysis they did not consider the contribution of the BL.
This result agrees with the changes in the hardness
ratio in the course of the bursts, as reported by other authors
(see, for instance, Paul et al. 1998). The average photon energy
of the PC must be higher in the irregular mode than in the regular
mode because pulses are more clearly detectable above 15 keV in
that state. During the irregular mode, when the PC energy
output was higher, the average count rate of the BL
decreased suggesting that some mechanism of energy transfer from
the SC to PC could be active in the source. All of these
considerations are useful to the spectral analysis of the
same data, which will be reported in a subsequent paper.

The results presented in this paper raise several interesting
problems to explain the physical processes producing this complex
behaviour of \grs. From a theoretical point of view, following the
first interpretation by Belloni et al. (1997a,b) and Taam, Chen
\& Swank (1997), the bursts of the $\rho$ class were modelled in
terms of thermal-viscous instabilities in the accretion disk.
Instabilities developing as series of pulses were investigated by
several authors well before the discovery of \grs. Taam \& Lin
(1984), Lasota \& Pelat (1991) computed light curves very similar
to those of \grss in the regular mode, identifying an SLT followed by a
narrow peak. From the numerical calculations of Taam \& Lin (1984),
however, there is no clear indication that the recurrence time
increases with the BL.
These authors also found an increase in the
pulse luminosity that is not observed in our results. About the
two above components, SC would correspond to the BL joint with the
pre-pulse emission, while PC is the overlying fast and prominent
pulse. Watarai \& Mineshige (2003) developed a model
for \grs, which includes an accretion disk around a 10 M$_{\odot}$
black hole, in which the viscosity stress has a functional
relationship with the integrated pressure that is more general than in
$\alpha$-disks.
They considered a factor $(p_{gas}/p)^{\mu}$ ($0<\mu<1$)
and calculated the light curves
for some values of $\mu$: the recurrence time decreases by a
factor of $\sim$2.5 for $\mu$ ranging from 0.1 to 0.2. However,
according to their plot, the recurrence time decreases when both the
DC level and pulse luminosity increase, which disagrees with our
findings.

In the framework of the Belloni et al. (1997a,b) model, the burst
sequence is interpreted as an emptying/refilling cycle of the
inner portion of the accretion disk. In this context, the rise
time - SLT in this work - is related to the viscous (disk-filling)
timescale, which is proportional to the 3.5-power of the inner disk radius,
while the pulse profile depends on the free-fall timescale, which
scales linearly with the radius. From this point of view, the
longer duration of SLT with respect to the {\it pulse}, is in
qualitative agreement with the model.

The $\rho$ class has been considered to be evidence of a
limit-cycle of the source defined in the parameter space
(Taam \& Lin 1984, Lasota \& Pelat 1991, Szuszkiewicz \& Miller 1998)
which encircles an instability region, such as the plane of
the disk temperature compared to the integrated surface density. Thus, the
two modes can provide information about changes occurring in the disk
structure. It is unclear why and how the transition between the
two modes occur, and the possible role of non-linear processes in
developing the instability. We note that \grss can remain in each
mode for relatively long time intervals, between many hours
and a few days, implying that the limit cycle continues for
several thousands of times. An additional unknown is the nature of the
complex limit-cycle behaviour exhibited by \grs. Limit cycles are
predicted by thermal-viscous instability and produce a
series of recurrent single bursts, such as those shown in the papers
quoted above. However, the pulse structure is far more complex,
and the observed number and duration of peaks vary between the
regular and the irregular mode. This implies minor limit cycles 
may exist within the major cycle. 
Such a possibility was discussed by Lasota \& Pellat (1991) in a 
different context, who also showed that an attractor existis along 
the stable branch of the
double-log plot of the disk temperature versus the integrated surface
density. It would be interesting to investigate wheter the increase in
multiplicity and the related change of mode are related or not to
the description of chaotic trajectories around the unstable branch
of an accretion disk.

\begin{acknowledgements}

The authors thank an anonymous referee for helpful refereal
of the paper, which helped to make it  clearer.
They are also grateful to the personnel of ASI Science Data
Center, particularly to M. Capalbi, for help in retrieving \sax
archive data. This work has been partially supported by research
funds of the Sapienza Universit\'a di Roma. The INAF Institutes
are financially supported by the Italian Space Agency (ASI) in the
framework of the contracts ASI-INAF I/023/05/0 (M.F.) and ASI-INAF
I/088/06/0 (T.B.).
\end{acknowledgements}

\appendix{

\section*{Appendix I - Data Reduction}

\begin{table*}
\caption{Log of the observations relative to the October 2000
pointing.}
\begin{center}
\begin{tabular}{lcrrc}
\hline
Obs. code & Tstart (UT) & \multicolumn{3}{c}{Exposure}   \\
                  & (dd-MM-yyyy hh:mm:ss) & MECS  & PDS  & name   \\
\hline
21226001   & 20-Oct-2000 21:26:55  & 32069 & 15343  & A \\
212260011  & 21-Oct-2000 21:19:43  & 37971 & 18538  & B \\ % (10045) D
212260012  & 22-Oct-2000 23:31:34  & 23042 & 11335  & C \\ % (10046) E
212260013  & 23-Oct-2000 15:58:56  & 39498 & 18938  & D \\ % (10050) F
212260014  & 24-Oct-2000 19:29:31  & 38873 & 18579  & E \\ % (10051) G
212260015  & 25-Oct-2000 23:16:11  & 41851 & 19834  & F \\ % (10052) H
212260016  & 27-Oct-2000 02:51:35  & 43547 & 20781  & G \\ % (10054) I
\hline
\end{tabular}
\end{center}
\end{table*}

The principal NFIs used in this observation were the Medium Energy
Concentrator Spectrometer (MECS) operating in the 1.3--10~keV
energy band (Boella et al. 1997b) and the Phoswich Detector System
(PDS) operating in the 15--300~keV energy band (Frontera et al.
1997). The source flux was also detected in the range 0.1-10 keV
by the Low Energy Concentrator Spectrometer (LECS) (Parmar et al.
1997) and by the HPGSPC (Manzo et al. 1997) in the range 6.0-30
keV, but these data were not used in this analysis because of
their low S/N. In particular, to avoid contamination of the
data by bright Earth radiation, the net LECS exposure was 
significantly lower than those of the MECS and the resulting light curve 
are of too short duration to be helpful for our analysis. 
Furthermore, the count rate below 1 keV was low
because of the high interstellar absorption towards GRS 1915+105.

Standard procedures and selection criteria were applied to the
data to avoid the South Atlantic Anomaly, solar, bright Earth, and
particle contamination, using the SAXDAS v. 2.0.0 package. The
images in the MECS contain a bright point-like source at a position
fully consistent with the precise radio coordinates. Events for
the time and spectral analysis were selected within circular
regions with a of radius $8'$, which contain about 95\% of the
point source signal in the MECS. As usual, the background was
estimated from high Galactic latitude `blank' fields imaged by
the same region of the detectors and its typical level corresponded
2.6$\times$10$^{-2}$ counts/s, much lower than the typical source
flux, even in the fainter states. PDS operated in the standard
``rocking mode" with two out of four units pointing at the source,
allowing for a contemporaneous background measurement. All light
curves were produced with a time binning of 0.5 s. The sum of the
durations of these time series provide the effective exposure times,
which are 256.8 ks for the MECS and 123.3 ks for the PDS,
respectively. Moreover, because of the Earth occultation and
occasional telemetry breakdown during the orbital motion of the
satellite, the resulting light curves, organised into a database,
consisted of a series of time segments separated by intervals of
variable length. A description of this database, including the
notation used for the time series, is given in the Appendix II,
and their durations and mean count rates, are reported in Table
A.2.

\section*{Appendix II - The light curve database and burst structure analysis}

We adopted an archive partition for the data series based on
the observation runs and the individual orbits.
Occasionally, some telemetry gaps of variable duration occurred.
To avoid having many very short series, the gaps shorter than 3
seconds were filled by interpolating nearby data. In the cases of
longer gaps, when interpolation could not be performed, the series
was further divided into more segments, whose durations are in the
range of between a few hundreds of seconds and about 3,000 seconds.

Light curves were divided into two main groups according to their
energy range (or the instrument): MECS (1.6--10 keV) and PDS
(15--100 keV) data, respectively. Every MECS data series was labelled
with a capital letter in alphabetical order to identify the
observations runs (see Table A.1), followed by a number
corresponding to the subsequent orbits in the run and, when
necessary, by a second lowercase letter to distinguish additional
time segments during an orbit caused by telemetry gaps (e.g., E9b
indicates the second segment of the ninth light curve of the fifth
run). We adopted the same notation for the observations codes and
orbit numbers for the PDS light curves, but because gaps are more
numerous and generally longer than in the MECS, these remain
unfilled and no distinction from the shortest segments was made. We
thus obtained 172 time series for the MECS, and many of them are
long enough for a good analysis, whereas a few others are very
short with only a small number of bursts and therefore are 
unsuitable for the Fourier analysis because of their coarse frequency
resolution.
Initial times (measured from the beginning of the observation
20 October 2000 at UT = 21$^{\rm h}$26$^{\rm m}$55$^{\rm s}$),
durations, and mean count rates are given in Table A.2.
Here we report, as an example, only the first ten lines of the table.
The complete Table A.2 is available at CDS.
The light curves of the \sax~ observation of October 2000
for the MECS (1.5--10 keV) and PDS (15--100 keV), used in the present analysis,
with the time resolution of 0.5 s can also be retrieved from CDS.

The second part of Table A.2 provides for each series the number of bursts of each
multiplicity, the average multiplicity, and the number of anomalous bursts
(see Sect 5.1).
Average multiplicities were found to be related to the regularity modes
and stability classes defined in Sect. 4:
the 36 series of stability class 2 have a mean multiplicity equal to 1.47,
with the standard deviation of 0.35, the 40 series of class 1
have a mean multiplicity of 1.50 ($\sigma$=0.36), whereas all the 18 series
of class 0 have an average multiplicity higher than 2 with a mean value
2.42 and $\sigma$=0.51.

\section*{Appendix III - Fourier periodograms classification}

We computed Fourier periodograms (hereafter FP) for the time segments
in the database of duration longer than 950 s, 
to estimate the most probable frequencies of burst repetition with
a sufficient resolution.
The number of useful time series for this analysis is 103, and they are
rather uniformly distributed over the entire observation interval.

On the basis of the number and height of the dominant peaks in FPs,
we classified the time segments into the following three types:
spectra of S type when a single peak is largely prominent over the
noise level or when the ratio of the powers of the highest
peak to the second one ($P_1/P_2$) is higher than a threshold
value, which we assumed equal to 3, and spectra of T type when two clearly 
separated peaks are apparent above the noise level and their power ratio is lower
than 3 and M type, when the high power features are more than two.

In Table A.3, FP results for the 103 considered series are reported:
for each segment we list the ratio $P_1/P_2$ of the powers of
the two highest peaks, the value of \trec~ corresponding to the maximum power
(or those of the two highest peaks for T series), or the centroid of the
period interval in which high signals are present (the number in parenthesis
measures the width of this interval, whose extreme values correspond to
a power equal to 1/3 of the maximum), together with the corresponding
time resolution of the FP that measures the uncertainty in \trec.
We note that the power ratio between the peaks was not computed in those
cases where the second peak was at the noise level.
The complete Table A.3 is available at CDS and in the on-line version.

This classification depends not only on the type of variations
exhibited by the source but also the time segmentation of series.
A different choice of the duration of the series could
modify the power distribution among the peaks in the FP, and the
peak height ratio may vary across the threshold. Nevertheless, it
provides a synthetic description of the mean properties of FPs. We
see that T types are the rarest, occurring only in the 14\% of the
series. We note also that M type spectra were preferentially found in
long segments: 25 of them of a total of 39 belong to series with
a duration longer than 2000 seconds. In contrast, S spectra are
more frequent in short series, and about 63\% were found in series
shorter than 2000 seconds.

The various spectral types are not randomly distributed in time, but some
of them are more frequent in one of the three intervals defined in
Sect. 3.
The S series are more numerous in the first and the third interval
(see definitions in Sect. 2),
the latter having rather long sequences of this type.
The D and M series appear in the first and more frequently in the second
interval: 14 of 26 series in the second interval
are of M type, and only seven are of S type.
Therefore, M series occur more frequently during the count rate bump.

We note that the values of \trec~ of the few T and M series present in
the last portion are characterised by small differences and quite
narrow ranges, respectively.
Moreover, the \trec~ found in the course of a sequence of S type
series, such as from E9a to E13 or F10a to F16a, does not remain constant
but changes from one to the subsequent stage by about 10\%.

\section*{Appendix IV - Wavelet transform and stability classification}

Standard wavelet analysis is based on the computation of wavelet power
spectra or {\it wavelet scalograms}, hereafter WS) defined as the
normalised square of the modulus of the wavelet transform

\begin{equation}\label{wav1}
W_{j,k} = \zeta  \left|w_{j}(\tau_{k})\right|^2 ~~,
\end{equation}

\noindent
where

\begin{equation}\label{wav2}
w_{j}(\tau_{k}) = \sqrt{\frac{\Delta t}{\tau_{k}}}~ \Sigma_i~ x_i~
\psi^*\Big( \frac{(i-j) \Delta t}{\tau_{k}}\Big)
\end{equation}

\noindent
and $\Delta t$ is the sampling time of the series $x_i$, $\psi^*$ the complex
conjugate of the wavelet function, $\zeta$ a normalisation factor and $\tau_k$
measures the timescale, usually starting from the sampling time
and increasing by doubling in terms of 2$^k$:

\begin{equation}\label{wavscale}
\tau_{k} = 2^{ak} (2 \Delta t)
\end{equation}

\noindent
where the constant $a$, which we assume to equal 0.05, changes the step size
of the scale sampling.
For a brief and practical introduction to wavelet analysis and its computation,
we refer to Torrence and Compo (1998).
There are many applications to astrophysical dataand a comprehensive
introduction to these methods can be found in Lachowicz and Czerny (2005).
In our analysis, we adopted the Morlet wavelet, which is a complex sinusoidal
waveform multiplied by a Gaussian bell profile

\begin{equation}\label{wavmor}
\psi(t) = \pi^{-1/4} e^{-i \omega t} e^{-t^2/2} ~~.
\end{equation}
\noindent
In the analysis, we used $\omega$=12, which is
well suited to studying the evolution of recurrence times in a burst series.

A quantitative evaluation of the stability of \trec~ in WSs can be
obtained by considering the changes of $T_{j,max}$, the timescale where the
maximum of the power $W_{j,k}$ is found for any time step $j$.
The plots of these maxima for the aforementined time series are shown on
right side of Fig. 10.
We see that for the S type A8b series the $T_{j,max}$ values are very close 
to that of the entire series at 49.0 s, and only occasionally varies
within the narrow interval between 47.5 and 51 s.
In the case of the T series F7, the variation range is larger and
the two most frequent timescales of the maxima coincide with the
peaks in the periodogram in Fig. 6.
The plot of the M type series E5 shows a variation range greater than
20 s in which rapid changes of $T_{j,max}$ are evident.
The ratio
\begin{equation}\label{wavsig}
R_w = \sigma_{T} / \langle T_{max} \rangle
\end{equation}
of the standard deviation of the $T_{J,max}$ to its mean value across
the series can be then taken as a measure of the stability of the
recurrence time.
The values of $\langle T_{max} \rangle$ and of $R_w$, given as
percentage, are also reported in Table A.3.

We introduced three stability classes based on the values of $R_w$:
spectra having $R_w < 3\%$ correspond to a high stability in the
recurrence of bursts and are indicated as class 2,
spectra with $3 < R_w < 8\%$ show a moderate stability
and are of class 1, and spectra with $R_w > 8\%$ are not
stable and belong to class 0.
These limits were chosen empirically from the inspection of the $R_w$
histogram and the series structure and were found to be satisfactory
for the goal of the series classification.
We found 36 series of class 2, 39 series of class 1, and 22 of class 0,
thus \grss was in a stable or quasi-stable state for the largest fraction
of the observation.
The complete classification, based on Fourier and wavelet analysis, is
given in the Table A.3.

The majority of S spectra, 30 out of 45, are of S2 type, 12 are S1, and 3
S0; we note also that 10 of the S1 spectra have $R_w$ smaller than 4.0,
indicating that their instability is small.
The three S0 series are D3a, D8b, and E3b, to which the S1 series D4b
must be added because it has $R_w$ slightly smaller than 8.
Two of these series belong to the time interval, defined in Sect. 4,
which corresponds to the increase in the count rate in the 6--10 keV
energy band, and the other two are just before it.
The majority of the 36 M spectra are of M0 or M1 type, and only 
four are of M2: i.e. B1b, F6, G2 and G5 and all these have peaks in the
periodograms of typical widths of only 4 seconds.
We note that there is essentially only one phase during which a high instability
was observed and it coincides with the second time fraction, when the mean
count rate in the 6--10 keV band was higher.

%%%%%%%%%%%%%%%%%%%%%%%%%%%%%%%%%%%%%%%%%%%%%%%%%%%%%%%%%%%%%%%%%%%%%%%%%%%

\begin{table*}
\caption{Initial times (starting from 20 October 2000 at UT=
21$^{\rm h}$26$^{\rm m}$55$^{\rm s}$), duration
and mean count rates in the MECS (1.6--10 keV), and PDS (15--100
keV) of all time series for the pointing of October 2000.
Multiplicity of bursts: for each series we report the number of
complete bursts, the numbers of bursts with multiplicities from
{\it p1} to {\it p4} and higher and the number of anomalous
bursts. The entire table is available at CDS or in the on-line version.}
\begin{center}
\begin{tabular}{lrrrrrrr|rrrrrrrr}
\hline
Series & Initial & Duration & C.R. & C.R. & C.R. & C.R. & C.R. &
 $N_b$ & {\it p1} & {\it p2} & {\it p3} & {\it p4} & {\it p$>$4} & $\langle p \rangle$ & An.\\
      & time   & MECS~  & MECS & MECS & MECS & MECS & PDS  &  &  &  &  &  &   &   & \\
      & s      & s      & ct/s & ct/s & ct/s & ct/s & ct/s &  &  &  &  &  &   &   & \\
\hline
%     &       &     &    &    &   &   &       &   & \\
A1a  &      12.5 &  737.5 & 203.5 & 43.6 &  96.3 & 58.5 & 39.6 &  15 & 13 &  2 & 0 & 0 & 0 & 1.13 & 0 \\ 
A1b  &    3268.5 & 1472.0 & 207.0 & 43.4 &  97.5 & 61.0 & 39.2 &  31 & 28 &  2 & 1 & 0 & 0 & 1.13 & 0 \\ 
A2a  &    6097.5 &  311.0 & 196.6 & 43.0 &  93.0 & 55.6 & 38.4 &   6 &  6 &  0 & 0 & 0 & 0 & 1.00 & 0 \\ 
A2b  &    8994.0 & 2191.0 & 203.1 & 43.5 &  95.3 & 58.0 & 39.6 &  43 & 39 &  4 & 0 & 0 & 0 & 1.09 & 0 \\ 
A3   &   14727.0 & 2509.5 & 203.9 & 42.8 &  96.3 & 59.5 & 39.2 &  53 & 40 & 11 & 2 & 0 & 0 & 1.28 & 0 \\ 
A4   &   20648.0 & 3152.0 & 198.0 & 41.8 &  92.8 & 58.1 & 36.5 &  66 & 32 & 33 & 1 & 0 & 0 & 1.53 & 0 \\ 
A5   &   26497.0 & 2809.5 & 193.4 & 40.3 &  90.1 & 56.7 & 35.1 &  57 & 31 & 24 & 2 & 0 & 0 & 1.49 & 0 \\ 
A6   &   31928.5 & 3108.0 & 191.8 & 40.8 &  89.9 & 56.3 & 35.4 &  64 & 34 & 29 & 1 & 0 & 0 & 1.48 & 0 \\ 
A7   &   37662.0 & 3156.5 & 199.2 & 42.7 &  93.8 & 57.1 & 38.4 &  62 & 51 & 11 & 0 & 0 & 0 & 1.18 & 0 \\ 
A8a  &   43402.0 &  577.5 & 192.5 & 41.6 &  90.7 & 55.6 & 36.0 &  10 &  9 &  1 & 0 & 0 & 0 & 1.10 & 0 \\ 
A8b  &   43996.0 & 2491.5 & 197.2 & 42.3 &  92.7 & 57.0 & 38.4 &  51 & 39 & 12 & 0 & 0 & 0 & 1.24 & 0 \\     
A9   &   49296.0 & 2952.5 & 206.1 & 42.9 &  94.0 & 57.0 & 41.9 &  56 & 44 & 12 & 0 & 0 & 0 & 1.21 & 0 \\     
A10a &   55310.0 &  654.5 & 214.4 & 46.3 & 101.8 & 62.3 & 40.9 &  12 & 10 &  2 & 0 & 0 & 0 & 1.17 & 0 \\     
A10b &   56028.5 & 1969.0 & 211.0 & 44.7 &  99.8 & 61.1 & 42.8 &  39 & 33 &  6 & 0 & 0 & 0 & 1.15 & 0 \\     
A11a &   61428.5 &  972.5 & 213.5 & 45.1 & 100.7 & 61.8 & 42.7 &  17 & 12 &  4 & 1 & 0 & 0 & 1.35 & 0 \\     
A11b &   62431.5 & 1288.5 & 212.7 & 45.1 &  99.1 & 60.7 & 43.6 &  24 & 21 &  3 & 0 & 0 & 0 & 1.12 & 0 \\     
A12  &   67565.5 &  973.5 & 212.6 & 45.2 & 101.1 & 62.0 &  --  &  19 & 19 &  0 & 0 & 0 & 0 & 1.00 & 0 \\    
A13a &   73698.5 &  781.5 & 209.2 & 44.2 &  99.2 & 61.7 &  --  &  16 & 16 &  0 & 0 & 0 & 0 & 1.00 & 0 \\         
A13b &   74524.5 &  153.5 & 196.8 & 42.4 &  93.3 & 58.0 &  --  &   2 &  2 &  0 & 0 & 0 & 0 & 1.00 & 0 \\    
A13c &   74692.5 &  255.5 & 202.6 & 44.0 &  96.3 & 58.0 &  --  &   4 &  4 &  0 & 0 & 0 & 0 & 1.00 & 0 \\    
A14  &   77898.0 &  479.0 & 210.5 & 44.5 & 100.7 & 60.9 &  --  &   9 &  9 &  0 & 0 & 0 & 0 & 1.00 & 0 \\    
B1a  &   85952.0 &  787.0 & 207.6 & 44.7 &  98.1 & 58.8 & 39.9 &  14 & 11 &  3 & 0 & 0 & 0 & 1.21 & 0 \\     
B1b  &   89265.0 & 1642.0 & 204.2 & 43.2 &  96.6 & 59.5 & 35.4 &  34 & 28 &  6 & 0 & 0 & 0 & 1.18 & 0 \\  
B2a  &   92067.0 &  325.5 & 202.3 & 42.9 &  94.2 & 57.1 & 35.4 &   6 &  6 &  0 & 0 & 0 & 0 & 1.00 & 0 \\  
B2b  &   94999.0 & 2197.0 & 207.6 & 43.9 &  97.9 & 60.3 & 40.6 &  45 & 38 &  7 & 0 & 0 & 0 & 1.16 & 0 \\  
B3   &  100740.0 & 2638.5 & 199.6 & 42.3 &  93.9 & 58.2 & 38.0 &  55 & 44 & 11 & 0 & 0 & 0 & 1.20 & 0 \\  
B4   &  107594.0 & 2008.0 & 207.6 & 44.2 &  96.9 & 59.3 & 40.5 &  38 & 28 & 10 & 0 & 0 & 0 & 1.26 & 0 \\  
B5   &  112199.0 & 3093.0 & 212.2 & 45.0 & 100.0 & 61.0 & 42.5 &  61 & 48 & 12 & 1 & 0 & 0 & 1.23 & 0 \\  
B6   &  117933.0 & 3146.5 & 208.3 & 44.0 &  98.4 & 60.4 & 40.7 &  65 & 51 & 14 & 0 & 0 & 0 & 1.22 & 0 \\  
B7   &  123666.5 & 3122.0 & 199.3 & 42.4 &  93.4 & 57.3 & 38.6 &  62 & 46 & 15 & 1 & 0 & 0 & 1.27 & 0 \\  
B8   &  129400.0 & 3057.5 & 203.3 & 43.1 &  95.8 & 59.3 & 39.2 &  61 & 50 &  9 & 2 & 0 & 0 & 1.21 & 0 \\  
B9a  &  135133.0 &  839.0 & 212.4 & 45.1 & 101.2 & 62.0 & 41.4 &  16 & 11 &  5 & 0 & 0 & 0 & 1.31 & 0 \\  
B9b  &  136123.5 & 2130.0 & 205.8 & 43.7 &  97.3 & 59.9 & 41.0 &  42 & 30 & 11 & 1 & 0 & 0 & 1.31 & 0 \\  
B10  &  141260.0 & 2685.0 & 210.1 & 44.7 &  98.5 & 60.7 & 41.6 &  52 & 36 & 16 & 0 & 0 & 0 & 1.31 & 0 \\  
B11a &  147399.0 &  808.5 & 213.4 & 44.0 & 102.5 & 62.5 & 42.4 &  15 & 12 &  3 & 0 & 0 & 0 & 1.20 & 0 \\
B11b &  148219.5 &  151.0 & 208.7 & 44.8 & 100.5 & 59.4 & 41.4 &   2 &  1 &  2 & 0 & 0 & 0 & 1.50 & 0 \\
B11c &  148566.0 &  783.5 & 210.7 & 44.2 & 100.5 & 61.7 & 41.8 &  15 & 14 &  1 & 0 & 0 & 0 & 1.07 & 0 \\
B12a &  153534.5 &  756.5 & 211.8 & 44.3 & 100.3 & 62.7 & 40.7 &  15 & 15 &  0 & 0 & 0 & 0 & 1.00 & 0 \\
B12b &  154298.0 & 1112.0 & 206.8 & 43.6 &  98.0 & 60.8 & 39.6 &  21 & 15 &  6 & 0 & 0 & 0 & 1.29 & 0 \\
B13a &  158068.0 &  138.5 & 207.9 & 43.2 &  99.2 & 61.1 & 36.2 &   2 &  2 &  0 & 0 & 0 & 0 & 1.00 & 0 \\  
B13b &  159666.0 & 1424.0 & 205.6 & 43.4 &  97.6 & 60.2 & 39.7 &  17 & 10 &  7 & 0 & 0 & 0 & 1.41 & 0 \\   
B14a &  163801.5 &  550.0 & 210.9 & 43.9 &  99.9 & 62.8 & 38.8 &  11 &  7 &  4 & 0 & 0 & 0 & 1.36 & 0 \\
B14b &  165933.0 &  918.5 & 206.6 & 43.0 &  97.7 & 61.6 & 37.1 &  19 & 12 &  7 & 0 & 0 & 0 & 1.37 & 0 \\
B15a &  169535.0 & 1077.5 & 210.3 & 42.7 &  98.8 & 64.3 & 37.5 &  20 &  3 & 15 & 2 & 0 & 0 & 1.95 & 0 \\
B15b &  171912.5 &  691.0 & 205.6 & 42.5 &  96.2 & 62.7 & 35.1 &  11 &  0 &  9 & 2 & 0 & 0 & 2.18 & 0 \\
B16a &  175268.5 & 1638.5 & 215.9 & 42.9 & 101.5 & 67.0 & 38.0 &  26 &  4 &  9 &12 & 1 & 0 & 2.38 & 1 \\
B16b &  178033.5 &  306.0 & 211.7 & 41.8 &  98.7 & 66.8 & 33.4 &   5 &  0 &  3 & 1 & 1 & 0 & 2.60 & 0 \\ 
\hline
\multicolumn{15}{c} { }
\end{tabular}
\end{center}
\end{table*}

\setcounter{table}{1}
\begin{table*}
\caption{Continued.}
\begin{center}
\begin{tabular}{lrrrrrrr|rrrrrrrr}
\hline
\hline
Series & Initial & Duration & C.R. & C.R. & C.R. & C.R. & C.R. &
 $N_b$ & {\it p1} & {\it p2} & {\it p3} & {\it p4} & {\it p$>$4} & $\langle p \rangle$ & An.\\
      & time   & MECS~  & MECS & MECS & MECS & MECS & PDS  &  &  &  &  &  &   &   & \\
      & s      & s      & ct/s & ct/s & ct/s & ct/s & ct/s &  &  &  &  &  &   &   & \\
\hline
C1   &  181003.0 & 2198.0 & 203.4 & 41.8 &  94.6 & 61.7 & 36.5 &  34 &  3 & 17 &12 & 1 & 1 & 2.41 & 3 \\   
C2   &  186736.5 & 2679.0 & 206.1 & 44.2 &  97.2 & 59.5 & 41.1 &  49 & 22 & 25 & 2 & 0 & 0 & 1.59 & 0 \\  
C3   &  192470.0 & 3103.5 & 206.6 & 43.6 &  97.4 & 60.5 & 39.2 &  61 & 41 & 20 & 0 & 0 & 0 & 1.33 & 0 \\  
C4   &  198203.5 & 3121.0 & 207.3 & 44.0 &  97.9 & 59.9 & 40.8 &  61 & 44 & 15 & 2 & 0 & 0 & 1.31 & 1 \\
C5   &  203937.0 & 3086.5 & 210.9 & 45.2 & 100.0 & 60.6 & 43.0 &  59 & 41 & 17 & 1 & 0 & 0 & 1.32 & 0 \\  
C6   &  209816.5 & 2942.0 & 211.6 & 43.9 &  98.5 & 61.4 & 40.2 &  59 & 30 & 28 & 1 & 0 & 0 & 1.51 & 0 \\  
C7   &  215404.0 & 3067.5 & 212.5 & 44.1 &  99.3 & 62.7 & 39.7 &  60 & 19 & 35 & 6 & 0 & 0 & 1.78 & 1 \\
C8a  &  221137.5 &  925.0 & 210.5 & 43.4 &  98.5 & 64.8 & 38.2 &  14 &  0 &  8 & 4 & 2 & 0 & 2.57 & 0 \\  
C8b  &  222298.5 & 1947.0 & 211.1 & 42.8 &  97.9 & 64.1 & 39.5 &  31 &  3 & 16 & 9 & 2 & 1 & 2.42 & 0 \\  
C9   &  227232.0 &  680.5 & 211.0 & 42.5 &  97.4 & 62.9 & 37.5 &  10 &  2 &  3 & 3 & 2 & 0 & 2.50 & 1 \\
D1a  &  239616.0 & 1816.5 & 210.6 & 43.7 &  98.1 & 63.1 & 40.6 &  26 &  1 & 14 & 7 & 4 & 0 & 2.54 & 3 \\
D1b  &  244071.5 &  107.5 & 221.4 & 45.1 & 103.2 & 68.8 & 36.4 &   1 &  0 &  1 & 0 & 0 & 0 & 2.00 & 0 \\  
D2a  &  245631.5 & 1535.0 & 208.6 & 42.8 &  96.6 & 62.4 & 40.0 &  22 &  1 & 13 & 5 & 2 & 1 & 2.50 & 1 \\ 
D2b  &  249805.0 &  560.5 & 201.0 & 42.5 &  94.3 & 59.8 & 41.7 &   7 &  1 &  4 & 1 & 1 & 0 & 2.29 & 2 \\
D3a  &  251755.0 & 1162.5 & 212.1 & 44.0 &  97.9 & 62.1 & 40.9 &  17 &  1 &  8 & 7 & 1 & 0 & 2.47 & 2 \\
D3b  &  255538.5 & 1102.5 & 207.4 & 43.7 &  97.5 & 60.8 & 38.7 &  20 &  9 & 11 & 0 & 0 & 0 & 1.55 & 0 \\  
D4a  &  257875.0 &  833.0 & 205.5 & 43.5 &  97.1 & 60.5 & 36.0 &  15 &  5 & 10 & 0 & 0 & 0 & 1.67 & 0 \\  
D4b  &  261271.5 & 1637.0 & 200.7 & 42.9 &  93.9 & 58.8 & 39.2 &  26 &  1 & 19 & 6 & 0 & 0 & 2.19 & 0 \\  
D5a  &  264109.5 &  287.5 & 226.7 & 46.5 & 107.8 & 67.7 & 43.4 &   5 &  4 &  1 & 0 & 0 & 0 & 1.20 & 0 \\  
D5b  &  267005.0 & 2219.0 & 215.3 & 45.3 & 101.2 & 63.2 & 40.9 &  43 & 29 & 13 & 1 & 0 & 0 & 1.35 & 0 \\  
D6   &  272748.0 & 2675.0 & 211.3 & 44.5 &  99.4 & 61.8 & 39.7 &  54 & 36 & 14 & 4 & 0 & 0 & 1.41 & 0 \\  
D7a  &  278741.0 & 1020.0 & 212.2 & 43.8 &  99.1 & 64.4 & 39.3 &  17 &  0 & 13 & 4 & 0 & 0 & 2.24 & 0 \\  
D7b  &  279523.0 & 1991.5 & 213.3 & 43.4 &  99.2 & 65.0 & 38.4 &  32 &  3 & 15 & 9 & 5 & 0 & 2.50 & 2 \\
D8a  &  284204.5 & 1262.0 & 223.9 & 44.3 & 104.1 & 69.1 & 39.9 &  20 &  1 & 12 & 7 & 0 & 0 & 2.30 & 0 \\  
D8b  &  285509.0 & 1825.0 & 224.9 & 43.8 & 103.5 & 70.0 & 37.4 &  27 &  2 & 11 & 8 & 4 & 2 & 2.74 & 4 \\
D9   &  289938.0 & 3117.5 & 228.7 & 44.9 & 105.9 & 71.1 & 39.7 &  50 &  5 & 22 &12 &10 & 1 & 2.60 & 6 \\
D10  &  295671.0 & 3129.5 & 227.8 & 44.5 & 105.1 & 71.1 & 39.0 &  45 &  0 & 12 &23 & 7 & 3 & 3.02 & 5 \\
D11  &  301404.5 & 2987.5 & 232.2 & 44.8 & 105.8 & 72.3 & 39.4 &  39 &  0 & 11 & 9 &15 & 4 & 3.31 & 4 \\
D12  &  307138.5 & 3158.0 & 225.8 & 44.7 & 104.7 & 70.7 & 40.2 &  40 &  1 & 10 &13 &10 & 6 & 3.25 & 6 \\
D13  &  313230.0 & 2730.5 & 216.5 & 43.5 & 100.6 & 66.3 & 39.1 &  40 &  6 & 15 &15 & 4 & 0 & 2.42 & 5 \\
D14a &  319365.0 &  831.5 & 220.8 & 44.4 & 102.7 & 69.0 & 41.0 &  25 &  0 & 11 & 8 & 4 & 2 & 2.88 & 3 \\
D14b &  320204.0 &  318.5 & 197.1 & 39.9 &  91.9 & 61.7 & 36.6 &   4 &  0 &  3 & 1 & 0 & 0 & 2.25 & 1 \\
D14c &  320534.5 &  600.0 & 210.9 & 42.3 &  98.1 & 65.9 & 39.1 &   7 &  0 &  2 & 3 & 2 & 0 & 3.00 & 1 \\
D14d &  321146.5 &  554.5 & 207.5 & 42.7 &  97.0 & 63.5 & 38.5 &   7 &  0 &  3 & 4 & 0 & 0 & 2.57 & 0 \\  
D15a &  325496.5 &  389.5 & 227.4 & 45.6 & 106.4 & 70.7 & 39.9 &   5 &  1 &  9 & 2 & 2 & 0 & 3.00 & 1 \\
D15b &  325895.0 &  586.0 & 220.4 & 44.2 & 103.1 & 68.4 & 38.6 &   8 &  0 &  2 & 3 & 2 & 1 & 3.25 & 2 \\
D15c &  326671.0 &  785.5 & 217.6 & 44.1 & 101.8 & 67.1 & 38.2 &  10 &  2 &  3 & 4 & 1 & 0 & 3.40 & 4 \\
D16a &  331624.5 &  741.5 & 223.7 & 45.1 & 104.4 & 69.5 & 41.2 &  10 &  1 &  3 & 4 & 2 & 0 & 2.70 & 2 \\
D16b &  332378.5 &  825.5 & 220.5 & 43.9 & 102.8 & 67.5 & 40.2 &  10 &  1 &  3 & 4 & 2 & 0 & 3.20 & 2 \\
D17a &  335807.5 &  585.0 & 221.3 & 45.2 & 102.7 & 68.6 & 40.6 &   8 &  1 &  5 & 2 & 0 & 0 & 2.12 & 0 \\  
D17b &  337760.0 &  760.5 & 220.5 & 45.2 & 103.1 & 67.4 & 40.7 &  10 &  1 &  6 & 2 & 0 & 1 & 2.40 & 0 \\
E1a  &  338580.5 &  328.5 & 222.8 & 45.0 & 104.2 & 68.9 & 42.6 &   4 &  0 &  2 & 1 & 1 & 0 & 2.75 & 0 \\     
E1b  &  341538.5 & 1129.5 & 230.3 & 45.8 & 105.8 & 70.0 & 41.1 &  17 &  1 & 10 & 4 & 2 & 0 & 2.41 & 0 \\  
E2a  &  343914.5 &  733.5 & 217.9 & 45.5 & 102.0 & 65.8 & 40.5 &  11 &  1 &  4 & 3 & 2 & 1 & 2.82 & 1 \\
E2b  &  347276.5 & 1705.5 & 228.8 & 45.9 & 106.0 & 69.5 & 41.6 &  25 &  1 & 13 & 7 & 2 & 2 & 2.64 & 1 \\
E3a  &  349997.0 &  359.5 & 212.9 & 43.1 & 100.8 & 64.5 & 41.0 &   5 &  0 &  2 & 1 & 2 & 0 & 3.00 & 0 \\  
E3b  &  353005.0 & 1193.0 & 222.4 & 44.6 & 103.4 & 67.9 & 40.8 &  16 &  1 &  8 & 5 & 2 & 0 & 2.50 & 1 \\
E3c  &  354268.5 &  954.5 & 210.8 & 43.2 &  98.1 & 63.8 & 37.0 &  14 &  1 &  5 & 5 & 2 & 1 & 2.79 & 3 \\
E4   &  358737.5 & 2666.5 & 221.3 & 44.1 & 102.8 & 67.8 & 38.9 &  39 &  4 & 18 &12 & 3 & 2 & 2.51 & 5 \\
E5   &  364470.5 & 3063.0 & 212.9 & 42.8 &  99.1 & 65.0 & 37.5 &  47 &  4 & 14 &21 & 7 & 1 & 2.72 & 6 \\
E6   &  370204.0 & 3025.5 & 204.7 & 42.6 &  94.4 & 59.8 & 39.6 &  42 &  3 & 19 &12 & 6 & 2 & 2.64 & 6 \\
\hline
\multicolumn{15}{c} { }
\end{tabular}
\end{center}
\end{table*}

\setcounter{table}{1}
\begin{table*}
\caption{Continued. }
\begin{center}
\begin{tabular}{lrrrrrrr|rrrrrrrr}
\hline
\hline
Series & Initial & Duration & C.R. & C.R. & C.R. & C.R. & C.R. &
 $N_b$ & {\it p1} & {\it p2} & {\it p3} & {\it p4} & {\it p$>$4} & $\langle p \rangle$ & An.\\
      & time   & MECS~  & MECS & MECS & MECS & MECS & PDS  &  &  &  &  &  &   &   & \\
      & s      & s      & ct/s & ct/s & ct/s & ct/s & ct/s &  &  &  &  &  &   &   & \\
\hline
E7   &  375937.0 & 3112.5 & 201.9 & 43.3 &  94.2 & 58.3 & 40.8 &  49 &  5 & 33 &11 & 0 & 0 & 2.12 & 1 \\
E8   &  381670.5 & 3091.0 & 212.3 & 45.6 & 100.3 & 60.4 & 42.7 &  59 & 40 & 19 & 0 & 0 & 0 & 1.32 & 0 \\  
E9a  &  387403.5 & 1751.5 & 211.5 & 45.3 &  99.5 & 59.5 & 42.9 &  33 & 25 &  7 & 1 & 0 & 0 & 1.27 & 1 \\
E9b  &  389326.5 & 1202.0 & 212.4 & 45.4 & 100.5 & 59.6 & 43.4 &  23 & 20 &  3 & 0 & 0 & 0 & 1.13 & 0 \\  
E10  &  393136.5 & 3062.0 & 215.8 & 46.4 & 102.5 & 61.1 & 44.2 &  55 & 45 & 10 & 0 & 0 & 0 & 1.18 & 0 \\  
E11a &  399203.0 &  977.5 & 222.4 & 47.9 & 105.8 & 63.5 & 45.4 &  16 & 14 &  1 & 1 & 0 & 0 & 1.19 & 0 \\  
E11b &  400209.0 & 1758.5 & 223.9 & 48.3 & 107.2 & 62.9 & 48.0 &  29 & 22 &  7 & 0 & 0 & 0 & 1.24 & 1 \\
E12  &  405336.0 & 2361.5 & 235.1 & 50.4 & 111.6 & 66.9 & 51.2 &  36 & 20 & 11 & 5 & 0 & 0 & 1.58 & 0 \\  
E13  &  411466.0 & 1767.5 & 241.2 & 51.0 & 113.4 & 68.3 & 51.4 &  27 & 18 &  6 & 3 & 0 & 0 & 1.44 & 1 \\
E14a &  416069.0 &   99.5 & 240.7 & 50.6 & 114.6 & 70.6 &  --  &   1 &  0 &  1 & 0 & 0 & 0 & 2.00 & 0 \\
E14b &  417590.5 &  753.0 & 246.9 & 52.3 & 118.4 & 71.1 & 54.3 &  10 &  7 &  2 & 1 & 0 & 0 & 1.40 & 0 \\
E14c &  418359.5 &  407.0 & 243.6 & 51.9 & 116.4 & 70.1 & 53.6 &   6 &  4 &  2 & 1 & 0 & 0 & 1.33 & 0 \\
E14d &  418778.5 &  389.0 & 238.7 & 50.9 & 114.9 & 68.1 & 52.5 &   6 &  4 &  2 & 1 & 0 & 0 & 1.33 & 0 \\
E15a &  421802.5 &  348.0 & 222.9 & 48.1 & 106.7 & 63.4 & 46.4 &   5 &  2 &  3 & 0 & 0 & 0 & 1.60 & 0 \\
E15b &  422161.5 &  195.5 & 225.5 & 48.2 & 108.8 & 64.4 & 47.0 &   2 &  1 &  1 & 0 & 0 & 0 & 1.50 & 0 \\  
E15c &  423712.5 & 1012.0 & 227.6 & 49.0 & 109.3 & 64.6 & 48.3 &  16 &  8 &  5 & 3 & 0 & 0 & 1.69 & 1 \\
E15d &  424744.0 &  141.5 & 212.5 & 46.3 & 100.4 & 61.1 & 45.1 &   2 &  1 &  1 & 0 & 0 & 0 & 1.50 & 0 \\  
E16a &  427536.0 &  371.0 & 217.1 & 46.9 & 103.9 & 61.8 & 48.5 &   4 &  2 &  1 & 1 & 0 & 0 & 1.75 & 0 \\
E16b &  429840.0 &  802.0 & 233.8 & 50.1 & 111.9 & 67.0 & 51.1 &  11 &  7 &  3 & 1 & 0 & 0 & 1.45 & 1 \\
E17a &  433280.0 & 1536.0 & 236.0 & 50.7 & 112.8 & 67.6 & 51.6 &  22 &  8 &  7 & 6 & 1 & 0 & 2.00 & 0 \\
E17b &  436128.0 &  254.5 & 238.3 & 50.9 & 113.7 & 68.7 & 50.4 &   2 &  1 &  1 & 0 & 0 & 0 & 1.50 & 0 \\
F1   &  439021.0 & 2194.0 & 243.0 & 51.6 & 115.9 & 70.1 & 52.5 &  31 & 21 &  7 & 3 & 0 & 0 & 1.42 & 0 \\    
F2   &  444735.5 & 2644.5 & 231.3 & 49.6 & 110.2 & 66.2 & 48.9 &  40 & 26 & 13 & 1 & 0 & 0 & 1.37 & 0 \\  
F3   &  450468.0 & 3020.0 & 231.9 & 50.1 & 110.3 & 65.9 & 49.5 &  45 & 29 & 11 & 5 & 0 & 0 & 1.47 & 2 \\
F4   &  456201.0 & 3102.0 & 241.8 & 51.6 & 114.1 & 68.6 & 51.6 &  47 & 37 &  4 & 6 & 0 & 0 & 1.34 & 0 \\  
F5a  &  461934.0 & 1346.0 & 244.7 & 52.7 & 117.7 & 70.0 & 53.5 &  19 & 12 &  1 & 5 & 0 & 1 & 1.79 & 0 \\  
F5b  &  463309.5 & 1707.5 & 244.4 & 52.4 & 117.1 & 69.9 & 53.4 &  26 & 14 &  8 & 3 & 1 & 0 & 1.65 & 0 \\  
F6   &  467752.0 & 2979.5 & 237.8 & 51.3 & 112.4 & 67.2 & 50.6 &  49 & 34 & 11 & 2 & 1 & 1 & 1.45 & 0 \\  
F7   &  473400.5 & 3099.0 & 234.3 & 50.1 & 110.8 & 66.3 & 49.4 &  51 & 33 & 15 & 2 & 1 & 0 & 1.43 & 0 \\  
F8   &  479133.5 & 2934.5 & 233.6 & 49.5 & 109.3 & 65.2 & 49.2 &  46 & 38 &  6 & 2 & 0 & 0 & 1.22 & 1 \\
F9   &  485200.0 & 2744.5 & 229.6 & 49.5 & 108.7 & 64.8 & 48.7 &  44 & 33 & 10 & 0 & 1 & 0 & 1.30 & 0 \\  
F10a &  491305.0 &  981.5 & 241.5 & 51.9 & 115.0 & 69.2 & 51.5 &  14 &  9 &  1 & 4 & 0 & 0 & 1.64 & 0 \\  
F10b &  492316.0 & 1357.5 & 242.7 & 51.5 & 114.9 & 69.1 & 53.4 &  19 & 14 &  3 & 1 & 1 & 0 & 1.42 & 1 \\
F11  &  497432.5 & 1994.0 & 239.4 & 50.7 & 113.6 & 68.0 & 50.3 &  31 & 16 & 13 & 2 & 0 & 0 & 1.55 & 1 \\
F12  &  503555.0 & 1552.5 & 237.6 & 51.1 & 113.2 & 68.4 & 50.8 &  23 & 14 &  9 & 0 & 0 & 0 & 1.39 & 0 \\  
F13a &  507808.0 &  615.0 & 235.8 & 50.7 & 113.0 & 67.2 & 49.3 &   7 &  4 &  2 & 1 & 0 & 0 & 1.57 & 0 \\  
F13b &  509728.0 & 1173.0 & 242.8 & 52.2 & 115.7 & 69.8 & 54.3 &  16 &  6 &  8 & 1 & 1 & 0 & 1.81 & 0 \\  
F14a &  513536.0 & 1187.0 & 240.3 & 51.9 & 114.8 & 68.5 & 54.8 &  16 & 12 &  2 & 2 & 0 & 0 & 1.37 & 0 \\  
F14b &  515800.0 &  879.5 & 242.1 & 52.1 & 115.5 & 69.5 & 49.1 &  12 &  6 &  2 & 3 & 1 & 0 & 1.92 & 0 \\  
F15a &  519264.0 & 1729.5 & 232.8 & 50.4 & 111.3 & 66.3 & 48.4 &  28 & 18 &  7 & 3 & 0 & 0 & 1.46 & 0 \\  
F15b &  521936.0 &  435.0 & 231.6 & 49.8 & 110.0 & 66.7 & 49.1 &   5 &  2 &  3 & 0 & 0 & 0 & 1.60 & 0 \\  
F16a &  524997.5 & 1461.0 & 249.5 & 54.1 & 119.1 & 71.0 & 55.4 &  19 &  8 &  6 & 4 & 1 & 0 & 1.89 & 0 \\
F16b &  526519.0 &  516.0 & 238.7 & 51.7 & 114.9 & 67.1 & 53.0 &   6 &  3 &  2 & 1 & 0 & 0 & 1.67 & 0 \\  
F17  &  530768.5 & 2578.0 & 244.3 & 52.2 & 116.4 & 70.3 & 51.9 &  36 & 15 & 17 & 3 & 1 & 0 & 1.72 & 1 \\   
F18a &  536463.5 &  689.0 & 242.3 & 51.9 & 115.8 & 69.5 & 52.9 &   8 &  5 &  3 & 0 & 0 & 0 & 1.37 & 0 \\  
F18b &  537166.0 &  675.0 & 244.0 & 52.1 & 116.5 & 70.3 & 53.3 &   9 &  4 &  4 & 1 & 0 & 0 & 1.67 & 0 \\
G1   &  537876.0 & 1583.0 & 246.4 & 52.5 & 119.3 & 72.5 & 53.6 &  21 & 13 &  7 & 0 & 1 & 0 & 1.48 & 0 \\     
G2a  &  542196.5 & 1040.0 & 240.5 & 54.7 & 123.1 & 75.5 & 54.5 &  14 & 10 &  4 & 0 & 0 & 0 & 1.29 & 0 \\
G2b  &  543248.5 & 1997.5 & 234.9 & 53.6 & 121.3 & 73.2 & 53.3 &  28 & 22 &  5 & 1 & 0 & 0 & 1.25 & 0 \\
G3   &  547929.0 & 3070.5 & 243.0 & 52.2 & 117.8 & 70.6 & 52.1 &  46 & 30 & 13 & 1 & 2 & 0 & 1.46 & 0 \\  
G4   &  553723.5 & 3026.0 & 234.1 & 50.6 & 113.1 & 68.0 & 48.4 &  50 & 39 &  9 & 2 & 0 & 0 & 1.26 & 0 \\  
G5   &  559395.0 & 3072.5 & 235.6 & 51.0 & 114.2 & 67.7 & 48.9 &  53 & 39 & 12 & 2 & 0 & 0 & 1.30 & 1 \\ 
\hline
\multicolumn{15}{c} { }
\end{tabular}
\end{center}
\end{table*}

\setcounter{table}{1}
\begin{table*}
\caption{Continued. }
\begin{center}
\begin{tabular}{lrrrrrrr|rrrrrrrr}
\hline
\hline
Series & Initial & Duration & C.R. & C.R. & C.R. & C.R. & C.R. &
 $N_b$ & {\it p1} & {\it p2} & {\it p3} & {\it p4} & {\it p$>$4} & $\langle p \rangle$ & An.\\
      & time   & MECS~  & MECS & MECS & MECS & MECS & PDS  &  &  &  &  &  &   &   & \\
      & s      & s      & ct/s & ct/s & ct/s & ct/s & ct/s &  &  &  &  &  &   &   & \\
\hline
G6a  &  565361.0 &  617.5 & 236.7 & 51.1 & 113.9 & 67.1 & 50.2 &   8 &  6 &  2 & 0 & 0 & 0 & 1.25 & 0 \\  
G6b  &  566000.5 & 2153.0 & 244.1 & 52.4 & 117.9 & 70.9 & 53.0 &  33 & 22 & 10 & 1 & 0 & 0 & 1.36 & 1 \\
G7   &  571145.0 & 2781.5 & 230.0 & 50.1 & 110.9 & 67.7 & 48.4 &  42 & 31 & 10 & 0 & 1 & 0 & 1.31 & 0 \\  
G8a  &  577273.0 &  981.5 & 227.9 & 49.2 & 110.2 & 65.1 & 48.8 &  15 &  9 &  5 & 1 & 0 & 0 & 1.47 & 0 \\  
G8b  &  578272.0 & 1405.5 & 231.5 & 49.6 & 110.3 & 66.3 & 49.3 &  22 & 18 &  2 & 2 & 0 & 0 & 1.27 & 0 \\  
G9   &  583495.5 & 1899.0 & 251.2 & 53.9 & 121.5 & 73.4 & 55.7 &  25 & 16 &  7 & 1 & 0 & 1 & 1.52 & 0 \\  
G10a &  589722.0 & 1363.0 & 249.5 & 53.6 & 120.5 & 72.8 & 55.1 &  16 &  8 &  6 & 1 & 1 & 0 & 1.69 & 0 \\  
G10b &  593168.5 &  424.0 & 252.1 & 53.8 & 122.5 & 74.0 & 53.7 &   4 &  1 &  3 & 0 & 0 & 0 & 1.75 & 0 \\  
G11a &  595637.0 & 1229.5 & 261.2 & 55.6 & 127.0 & 77.9 & 57.6 &  14 &  7 &  2 & 4 & 1 & 0 & 1.93 & 0 \\  
G11b &  599526.0 & 1212.0 & 252.3 & 53.7 & 122.1 & 73.8 & 54.1 &  15 & 11 &  3 & 1 & 0 & 0 & 1.33 & 0 \\  
G12a &  601760.0 &  846.0 & 252.9 & 54.4 & 123.0 & 75.8 & 55.0 &  10 &  4 &  5 & 0 & 1 & 0 & 1.80 & 0 \\  
G12b &  605264.0 &  904.5 & 252.1 & 54.1 & 121.6 & 73.4 & 56.1 &   9 &  4 &  3 & 2 & 0 & 0 & 1.78 & 0 \\  
G12c &  606320.0 &  677.0 & 248.1 & 55.0 & 126.1 & 77.1 & 56.6 &   6 &  4 &  2 & 0 & 0 & 0 & 1.33 & 0 \\  
G13a &  607904.0 &  403.5 & 265.6 & 55.0 & 126.6 & 78.5 & 56.7 &   4 &  3 &  1 & 0 & 0 & 0 & 1.25 & 0 \\
G13b &  610992.0 & 1264.5 & 262.3 & 55.5 & 125.1 & 76.2 & 59.5 &  11 &  8 &  3 & 0 & 0 & 0 & 1.27 & 0 \\
G14  &  617328.0 & 1949.5 & 295.6 & 60.1 & 140.6 & 88.9 & 58.4 &  29 & 10 & 16 & 2 & 1 & 0 & 1.79 & 1 \\  
G15a &  622457.0 & 1068.0 & 273.2 & 57.3 & 130.2 & 80.1 & 59.8 &  11 &  5 &  3 & 3 & 0 & 0 & 1.82 & 0 \\
G15b &  623611.5 & 1805.0 & 272.9 & 57.1 & 130.1 & 80.0 & 59.8 &  20 & 13 &  7 & 0 & 0 & 0 & 1.35 & 0 \\
G16a &  628188.5 &  953.5 & 263.7 & 55.8 & 125.3 & 77.3 & 61.4 &   9 &  6 &  2 & 1 & 0 & 0 & 1.44 & 0 \\
G16b &  629170.5 & 2070.0 & 269.4 & 56.4 & 128.6 & 78.9 & 59.2 &  21 &  9 &  8 & 3 & 1 & 0 & 1.81 & 0 \\
G17a &  633921.5 & 1267.5 & 273.8 & 56.8 & 131.4 & 80.0 & 59.7 &  15 & 11 &  3 & 1 & 0 & 0 & 1.33 & 0 \\
G17b &  635224.5 & 1749.0 & 268.9 & 56.2 & 128.2 & 78.9 & 58.7 &  22 & 11 &  8 & 3 & 0 & 0 & 1.64 & 0 \\
\hline
\multicolumn{15}{c} { }
\end{tabular}
\end{center}
\end{table*}

\setcounter{table}{2}
\begin{table*}
\caption{Results of the Fourier and wavelet spectral analysis and the
classification of the considered {\bf 103} MECS time series.
For each series, we indicate the name code, the $P1/P2$ power ratio of two
highest peaks, the recurrence time of the unique (for the S series) and
of the two (for the T series) dominant peaks or the centroid and the
interval width (for the M series), the temporal resolution at the peak,
the timescale of the maximum in the wavelet spectra, the ratio $R_w$,
and the series classification.
The entire table is available at CDS or in the on-line version.}
\begin{center}
\begin{tabular}{lclcrrc}
\hline
Series & P1/P2 & \trec & $\Delta t$& $\langle T_{max} \rangle$ & $R_w$ & Type \\
       &    & ~~~s  & s      & s~~~                     &    &  \\
\hline
%     &   & &   &         &   &   & &  & \\
A1b  & ---  & 44.0       & 1.3 & 45.4 & 2.00 & S2 \\
A2b  & 3.70 & 49.0       & 1.1 & 49.1 & 2.56 & S2 \\
A3   & 3.98 & 46.0       & 0.8 & 46.6 & 1.82 & S2 \\
A4   & 1.90 & 46.8 (4.4) & 0.7 & 47.1 & 3.56 & M1 \\
A5   & 1.15 & 46.5 (9.3) & 0.7 & 48.0 & 8.46 & M0 \\
A6   & 1.05 & 47.0 (5.0) & 0.7 & 47.7 & 4.61 & M1 \\
A7   & 1.09 & 49.4 (4.6) & 0.8 & 49.6 & 4.16 & M1 \\
A8b  & ---  & 48.8       & 1.0 & 48.7 & 2.24 & S2 \\
A9   & 3.58 & 51.4       & 0.9 & 51.3 & 2.70 & S2 \\
A10b & 3.48 & 49.0       & 1.3 & 48.8 & 2.00 & S2 \\
A11a & 2.46 & 47.5; 46.3 & 0.8 & 47.3 & 3.28 & T1 \\
B4   & 2.32 & 47.0; 50.0 & 0.7 & 47.7 & 2.84 & T2 \\
B7   & 1.52 & 49.0 (4.6) & 0.8 & 49.2 & 5.07 & M1 \\
B8   & 2.07 & 48.5 (6.0) & 0.8 & 49.5 & 5.99 & M1 \\
B9b  & 2.58 & 45.0; 49.0 & 1.1 & 49.4 & 3.85 & T1 \\
B10  & 2.03 & 49.7 (7.0) & 1.0 & 51.0 & 5.58 & M1 \\
B12b & 1.89 & 48.1 (10.2)& 2.0 & 51.0 & 16.7 & M0 \\  
B13b & 1.71 & 50.5 (14.2)& 2.4 & 52.7 & 12.7 & M0 \\  
B15a & 1.00 & 49.4; 53.0 & 2.2 & 51.8 & 13.7 & T0 \\  
B16a & 1.04 & 61.0 (7.1) & 2.1 & 62.1 & 9.50 & M0 \\  
C1   & 1.59 & 65.4 (16.0)& 2.2 & 69.7 & 15.6 & M0 \\
C2   & 2.14 & 56.0; 50.0 & 1.2 & 54.0 & 6.09 & T1 \\
C3   & 1.14 & 49.0 (7.0) & 0.8 & 49.5 & 4.22 & M1 \\
C4   & 5.82 & 49.5       & 0.8 & 50.9 & 4.66 & S1 \\
C5   & 2.59 & 49.4 (4.0) & 0.8 & 51.4 & 7.31 & M1 \\
C6   & 1.51 & 48.2 (8.4) & 0.9 & 48.5 & 6.93 & M1 \\ 
C7   & 1.29 & 50.7 (9.1) & 0.9 & 50.8 & 6.40 & M1 \\
C8b  & 2.16 & 60.9 (16.0)& 2.1 & 62.8 & 13.4 & M0 \\    
D1a  & 4.33 & 66.0       & 2.4 & 67.4 & 3.74 & S1 \\
D2a  & 1.12 & 61.5 (11.0)& 2.4 & 62.5 & 10.2 & M0 \\
D3a  & 3.55 & 59.0       & 3.0 & 68.8 & 14.9 & S0 \\
D3b  & 1.71 & 48.0; 52.0 & 2.1 & 51.6 & 7.74 & T1 \\
D4b  & 3.43 & 57.0       & 2.0 & 61.1 & 7.68 & S1 \\
D5b  & 1.77 & 50.6 (3.0) & 1.1 & 50.9 & 4.76 & M1 \\
D6   & 2.68 & 48.6; 50.5 & 0.9 & 49.1 & 5.06 & T1 \\
D7a  & 5.12 & 55.0       & 3.0 & 56.4 & 2.65 & S2 \\
D7b  & 1.66 & 61.0; 67.0 & 1.9 & 63.3 & 13.4 & T0 \\
D8a  & 2.96 & 59.0; 63.0 & 2.7 & 60.3 & 3.43 & T1 \\
D8b  & 3.16 & 64.0       & 2.2 & 67.8 & 16.4 & S0 \\
D9   & 1.17 & 64.9 (16.0)& 1.5 & 63.0 & 10.3 & M0 \\
D10  & 1.43 & 69.1 (19.0)& 1.4 & 68.9 & 10.2 & M0 \\
D11  & 2.53 & 68.1 (4.0) & 1.6 & 95.2 & 55.6 & M0 \\
D12  & 1.02 & 77.5 (16.0)& 1.7 & 114.3 & 59.4 & M0 \\ 
D13  & 1.36 & 71.2 (34.0)& 1.8 & 67.9 & 14.2 & M0 \\
E1b  & 1.73 & 69.2 (20.0)& 4.7 & 68.5 & 9.01 & M0 \\     
E2b  & 1.68 & 63.2 (15.0)& 2.7 & 67.8 & 9.44 & M0 \\     
E3b  & 3.88 & 68.0       & 3.9 & 65.7 & 11.4 & S0 \\     
E3c  & 5.53 & 63.0       & 4.2 & 66.8 & 2.67 & S2 \\   
E4   & 2.26 & 58.0; 71.0 & 1.9 & 70.7 & 7.56 & T1 \\   
E5   & 1.35 & 67.0 (24.0)& 1.7 & 68.2 & 9.54 & M0 \\  
\hline
\multicolumn{7}{c} { }
\end{tabular}
\end{center}
\end{table*}

\setcounter{table}{2}
\begin{table*}
\caption{Continued.}
\begin{center}
\begin{tabular}{lclcrrc}
\hline
Series & P1/P2 & \trec & $\Delta t$& $\langle T_{max} \rangle$ & $R_w$ & Type \\
       &    & ~~~s  & s      & s~~~                     &    &  \\
\hline
E6   & 1.04 & 68.2 (16.0)& 1.5 & 72.1 & 11.5 & M0 \\  
E7   & 2.87 & 64.9 (12.0)& 1.5 & 63.2 & 9.25 & M0 \\ 
E8   & 2.99 & 51.0; 52.0 & 0.8 & 51.4 & 2.71 & T2 \\ 
E9a  & 6.24 & 51.0       & 1.5 & 51.6 & 2.71 & S2 \\ 
E9b  & ---  & 50.0       & 2.1 & 51.7 & 2.17 & S2 \\ 
E10  & 6.99 & 53.0       & 0.9 & 54.0 & 2.72 & S2 \\ 
E11a & 5.14 & 54.0       & 3.0 & 55.0 & 1.71 & S2 \\ 
E11b & 4.58 & 57.0       & 1.8 & 57.5 & 1.97 & S2 \\ 
E12  & ---  & 63.0       & 1.7 & 63.7 & 1.68 & S2 \\ 
E13  & ---  & 60.6       & 2.0 & 61.8 & 3.37 & S1 \\ 
E15c & ---  & 57.8       & 3.3 & 67.6 & 2.43 & S2 \\
E17a & ---  & 65.5       & 2.8 & 67.9 & 2.55 & S2 \\
F1   & 1.01 & 60.8 (5.0) & 1.9 & 68.2 & 3.47 & M1 \\ 
F2   & 7.00 & 64.0       & 1.5 & 64.4 & 2.30 & S2 \\ 
F3   & 7.13 & 66.0       & 1.4 & 66.7 & 3.10 & S1 \\ 
F4   & ---  & 65.0       & 1.4 & 64.8 & 1.82 & S2 \\ 
F5a  & 1.39 & 62.7 (6.0) & 1.4 & 65.0 & 3.65 & M1 \\ 
F5b  & 1.10 & 62.1 (6.0) & 2.1 & 64.0 & 4.24 & M1 \\
F6   & 1.33 & 59.2 (4.0) & 1.2 & 59.5 & 2.50 & M2 \\   
F7   & 1.28 & 58.0; 60.0 & 1.1 & 59.6 & 2.76 & T2 \\   
F8   & ---  & 62.0       & 1.3 & 62.1 & 1.97 & S2 \\   
F9   & 1.26 & 61.1 (5.0) & 1.4 & 61.8 & 3.80 & M1 \\   
F10a & ---  & 65.0       & 4.4 & 68.0 & 2.53 & S2 \\    
F10b & 8.05 & 66.0       & 3.2 & 67.4 & 2.93 & S2 \\   
F11  & 7.56 & 60.0       & 1.8 & 61.9 & 1.39 & S2 \\   
F12  & ---  & 65.7       & 2.7 & 67.0 & 1.99 & S2 \\
F13b & ---  & 66.0       & 3.8 & 69.1 & 0.00 & S2 \\    
F14a & 5.84 & 68.0       & 3.9 & 71.0 & 2.58 & S2 \\    
F15a & 3.33 & 63.0       & 2.3 & 63.7 & 3.10 & S1 \\   
F16a & ---  & 69.6       & 3.3 & 72.0 & 3.25 & S1 \\  
F17  & 1.20 & 67.0; 69.7 & 1.7 & 67.7 & 4.89 & T1 \\  
G1   & 7.37 & 68.6 (4.0) & 1.6 & 70.3 & 2.56 & M2 \\   
G2b  & ---  & 67.7       & 2.3 & 69.3 & 1.64 & S2 \\
G3   & 1.69 & 65.1 (5.0) & 1.3 & 65.7 & 3.13 & M1 \\   
G4   & 5.13 & 58.0       & 1.1 & 59.5 & 3.47 & S1 \\   
G5   & 1.46 & 58.5 (4.0) & 1.1 & 59.0 & 2.91 & M2 \\   
G6b  & ---  & 64.0       & 1.9 & 65.3 & 2.88 & S2 \\   
G7   & 1.41 & 61.9 (6.0) & 1.4 & 64.1 & 4.55 & M1 \\   
G8a  & ---  & 61.0       & 3.8 & 63.1 & 2.36 & S2 \\   
G8b  & 3.79 & 62.0       & 2.8 & 64.3 & 3.34 & S1 \\   
G9   & ---  & 73.0       & 2.8 & 74.9 & 1.69 & S2 \\   
G10a & 8.32 & 76.0       & 4.2 & 78.7 & 2.51 & S2 \\   
G11a & 3.11 & 82.0       & 5.5 & 84.4 & 2.08 & S2 \\   
G11b & 1.91 & 74.0 (13.0)& 4.7 & 75.0 & 6.16 & M1 \\        
\hline
\multicolumn{7}{c} { }
\end{tabular}
\end{center}
\end{table*}


\begin{thebibliography}{ }
\small

\bibitem[Belloni et al. 1997a]{bel_97a}
Belloni, T. et al., 1997a, ApJ 488, L109

\bibitem[Belloni et al. 1997b]{bel_97b}
Belloni, T. et al., 1997b, ApJ 479, L145

\bibitem [Belloni et al. 2000]{bel_00}
Belloni, T., Klein-Wolt, M., M\'endez, M. et al., 2000,
A\&A 355, 271

\bibitem [Belloni 2001]{bel_01}
Belloni, T., M\'endez, M., S\'anchez-Fernandez C., 2001,
A\&A 372, 551

\bibitem[Boella et al. 1997a]{boe_1}
Boella, G., Butler, R.C., Perola, G.C. et al., 1997a,
A\&AS 122, 299

\bibitem[Boella et al. 1997b]{boe_2}
Boella, G., Chiappetti, L., Conti, G. et al., 1997b,
A\&AS 122, 327

\bibitem[Castro-Tirado A.J. et al. 1992]{cas_1}
Castro-Tirado, A.J., Brandt, S., Lund, N., 1992,
IAUC N. 5590

\bibitem [FenderBelloni 2004]{fen_1}
Fender, R., Belloni, T., 2004,
ARAA 42, 317

\bibitem [Feroci et al. 1999]{fer_1}
Feroci, M., Matt, G., Pooley G. et al., 1999,
A\&A 351, 985

\bibitem [Feroci et al. 2001]{fer_2}
Feroci, M., Casella P., Costa E. et al., 2001,
Proc. 3rd Microquasar Workshop,
ApSS 276(suppl), 15

\bibitem [Frontera et al. 1997]{fro_1}
Frontera, F., Costa, E., Dal Fiume, D. et al., 1997,
A\&AS 122, 357

\bibitem [Gimenez et al. 2001]{gim_1}
Gim\'enez de Castro, C.G., Raulin, J.-P., Mandrini, C.H. et al., 2001,
A\&A 366, 317

\bibitem [Greiner et al. 2001]{gre_1}
Greiner, J., Cuby, J.G., McCaughrean, M.J., 2001,
Nature 414, 522

\bibitem [Hannikainen et al. 2003]{han_1}
Hannikainen, D.C., Vilhu, O., Rodriguez, J. et al., 2003,
A\&A 411, L415

\bibitem [Hannikainen et al. 2005]{han_2}
Hannikainen, D.C., Rodriguez, J., Vilhu, O. et al., 2005,
A\&A 435, 995

\bibitem [Harlaftisgreiner 2004]{har_1}
Harlaftis E.T., Grenier J., 2004,
A\&A 414, L3

\bibitem [Kleinwolt et al. 2004]{Kw_1}
Klein-Wolt, R.P., Fender, R.P., Pooley, G.G. et al., 2002,
MNRAS 331, 745

\bibitem[LachCze 2005]{Lac_1}
Lachowicz, P., Czerny, B., 2005,
MNRAS 361, 645

\bibitem[Lasotapelat 1991]{Las_1}
Lasota, J.P., Pelat, D., 1991,
A\&A 249, 574

\bibitem[McClintock \& Remillard 2006]{mcclintock}
McClintock, J.E. \& Remillard, R.A., in "Compact Stellar X-rays
Sources", Eds W. Lewin \& M. van der Kils, Cambridge University
Press, 2006, p. 157

\bibitem[Mirabel Rodriguez 1994]{mir_1}
Mirabel, I.F., Rodriguez, L.F., 1994,
Nature 371, 46

\bibitem[Misra et al. 2006]{misr_02}
Misra, R., Harikrishnan, K.P. et al., 2006,
APJ 643, 1114

\bibitem[Morgan et al. 1997]{morg_97}
Morgan, E.H., Remillard, R.A., Grenier J., 1997,
APJ 482, 993

\bibitem[Nayakshin et al. 2000]{nay_1}
Nayakshin, S., Rappaport, S., Melia, F., 2000,
ApJ 535, 798

\bibitem[NaikRaoChak 2002]{nai_1}
Naik, S., Rao, A.R, Chakrabarti S., 2002,
J. Astron. Astrophys. 23, 213

\bibitem[Naik et al. 2002]{nai_2}
Naik, S., Agrawal, P.C., Rao, A.R, Paul, B., 2002,
MNRAS 330, 487

\bibitem[Paul et al. 1998]{pau_1}
Paul, B., Agrawal, P.C., Rao, A.R. et al., 1998,
ApJ 492, L63

\bibitem[Rodriguez et al. 2008]{rod_1}
Rodriguez J., Hannikainen D.C., Shaw S.E. et al., 2008,
ApJ 675, 1436

\bibitem[Szuszkiewicz Miller 1998]{szumil_1}
Szuszkiewicz, E., Miller, J.C., 1998,
MNRAS 298, 888

\bibitem[Taam Lin 1984]{taa_1}
Taam, R.E., Lin, D.N.C., 1984,
ApJ 287, 761

\bibitem[Taam et al. 1997]{taa_2}
Taam, R.E., Chen, X., Swank, J.H., 1997,
ApJ 485, L83

\bibitem[TorrCompo 1998]{tor_1}
Torrence, C., Compo, G.P., 1998,
BAMS 79, 61

\bibitem[Ueda 2002]{ued_1}
Ueda, Y., Yamaoka, K., S\'anchez-Fern\'andez, C. et al., 2002,
ApJ 571, 918

\bibitem[Vilhu 1998]{vil_1}
Vilhu, O., Nevalainen, J., 1998,
ApJ 508, L85

\bibitem[Watarai 2003]{vat_1}
Watarai, K.-Y., Mishinege, S., 2003,
ApJ 596, 421

\end{thebibliography}
\end{document}